\documentclass[aps,prb,amsmath,amssymb,floatfix,twocolumn]{revtex4-1}
\usepackage{graphicx}
\usepackage{subfigure}
\usepackage{color}
\usepackage{hyperref}

\begin{document}

\title{Axial anomaly and longitudinal magnetoresistance of a generic three dimensional metal}
\author{Pallab Goswami}
\affiliation{Condensed Matter Theory Center and Joint Quantum Institute, Department of Physics, University of Maryland, College Park, Maryland 20742-4111, USA}
\author{J. H. Pixley}
\affiliation{Condensed Matter Theory Center and Joint Quantum Institute, Department of Physics, University of Maryland, College Park, Maryland 20742-4111, USA}
\author{S. Das Sarma}
\affiliation{Condensed Matter Theory Center and Joint Quantum Institute, Department of Physics, University of Maryland, College Park, Maryland 20742-4111, USA}

\begin{abstract}
We show that the emergence of the axial anomaly is a universal phenomenon for a generic three dimensional metal in the presence of parallel electric ($E$) and magnetic ($B$) fields. In contrast to the expectations of the classical theory of magnetotransport, this intrinsically quantum mechanical phenomenon gives rise to the longitudinal magnetoresistance for any three dimensional metal. However, the emergence of the axial anomaly does not guarantee the existence of negative longitudinal magnetoresistance. We show this through an explicit calculation of the longitudinal magnetoconductivity in the quantum limit using the Boltzmann equation, for both short-range neutral and long-range ionic impurity scattering processes. We demonstrate that the ionic scattering contributes a large positive magnetoconductivity $\propto B^2$ in the quantum limit, which can cause a strong negative magnetoresistance for any three dimensional or quasi-two dimensional metal. In contrast, the finite range neutral impurities and zero range point impurities can lead to both positive and negative longitudinal magnetoresistance depending on the underlying band structure. In the presence of both neutral and ionic impurities, the longitudinal magnetoresistance of a generic metal in the quantum limit initially becomes negative, and ultimately becomes positive after passing through a minimum. We discuss in detail the qualitative agreement between our theory and recent observations of negative longitudinal magnetoresistance in Weyl semimetals TaAs and TaP, Dirac semimetals Na$_3$Bi, Bi$_{1-x}$Sb$_x$, and ZrTe$_5$, and quasi-two dimensional metals PdCoO$_2$, $\alpha$-(BEDT-TTF)$_2$I$_3$ which do not possess any bulk three dimensional Dirac or Weyl quasiparticles.
\end{abstract}

\maketitle
\section{Introduction} \label{sec:Intro}
The axial anomaly plays a profound role in the construction of effective relativistic field theories in odd spatial dimensions, involving massless Dirac and Weyl fermions~\cite{peskin,fujikawa}. When massless Dirac fermions in one spatial dimension are coupled to an external electric field ($E$), the separate number conservation laws of the right and the left handed fermions are broken by a quantum mechanical effect~\cite{peskin,fujikawa}, which is described by
\begin{equation}
\partial_{\mu}(j_{\mu,R}- j_{\mu,L})=\partial_{\mu} j_{\mu,5}=\frac{eN_fE}{\pi \hbar}, \label{ABJ1}
\end{equation} where $N_f$ is the number of flavors, $e$ is the electric charge, and $j_{\mu,R/L}$ describe the charge and the current operators of the chiral fermions. Similarly, in three spatial dimensions, in the presence of parallel electric ($\mathbf{E}$) and magnetic ($\mathbf{B}$) fields, the anomalous violation of the axial current conservation law is described by the celebrated Adler-Bell-Jackiw (ABJ) anomaly equation~\cite{adler,bell}
\begin{equation}
\partial_\mu j_{\mu,5}=\frac{e^2 N_f}{2 \pi^2 \hbar} \; \mathbf{E}\cdot\mathbf{B}. \label{ABJ2}
\end{equation}  The emergence of the axial anomaly is intimately tied to the topological properties of the Dirac operator with unbounded linear dispersion. It is related to the spectral flow of the right and left handed states in the presence of a topologically nontrivial configuration of the gauge field: one sinks below zero energy and simultaneously the other one rises above zero energy through an infinite Dirac sea. For constructing chiral gauge theories such as the electroweak theory and the standard model, it is important to cancel the anomaly contributions of different species or flavors of chiral fermions~\cite{peskin}. However, it is not exactly clear how the notion of the axial anomaly can have any impact on the physical properties of a condensed matter system endowed with a \emph{bounded} dispersion relation, since the infinite nature of the Dirac sea is crucial to the existence of the ABJ anomaly in quantum field theories.

Nevertheless in a seminal paper, Nielsen and Ninomiya~\cite{nielsen} first proposed that the axial anomaly can give rise to a very large longitudinal magnetoconductivity (LMC) or a negative longitudinal magnetoresistance (LMR) for certain gapless semiconductors, which possess a linear touching of conduction and valence bands at isolated points of the Brillouin zone. Depending on whether the bands are nondegenerate or two-fold degenerate, the gapless semiconductor is respectively known as a Weyl semimetal (WSM) or a Dirac semimetal (DSM). The recent discovery of a topological DSM ~\cite{Dai1,Dai2}($N_f=2$) in Cd$_3$As$_2$~\cite{Zahid1,Cava1,ZX1}, Na$_3$Bi~\cite{ZX2,Zahid2}, and the search for a WSM in the magnetically ordered (all in-all out) 227 pyrochlore iridates~\cite{vishwanath}, have rekindled the interest of the condensed matter physics community in experimental verification of the axial anomaly. Very recently a WSM phase with an even number of left and right handed pairs has been discovered in noncentrosymmetric TaAs~\cite{TaAs1,TaAs2,TaAs3,TaAs4}. The ideas are being put forward that the verification of the axial anomaly induced negative LMR can serve as a diagnostic tool for the existence of WSM and DSM~\cite{aji,son,vishwanathsid,shovkovy,burkov1}. A DSM phase with an odd number of flavors also arises at the quantum phase transition transition between a strong $Z_2$ topological insulator and a trivial band insulator in various materials~\cite{Goswami1} such as Bi$_{1-x}$Sb$_x$~\cite{Lenoir,Goswami2,Teo}, BiTl(S$_{1-\delta}$Se$_\delta$)$_2$~\cite{Xu1,Sato}, (Bi$_{1-x}$In$_x$)$_2$Se$_3$~\cite{Brahlek,Liu}. Similarly, a DSM with even number of cones are known to appear at the transition between a crystalline topological insulator and a trivial insulator in Pb$_{1-x}$Sn$_x$Te~\cite{Nimtz,Zahid3,Cava2}. A different kind of gapless fermion, known as a massless Kane fermion, is realized in Hg$_{1-x}$Cd$_x$Te, at the band inversion transition between a parabolic semimetal with quadratic band touching (Hg rich side) and a semiconductor (Cd rich side)~\cite{Nimtz,Orlita}. The gapless Kane fermion is constructed out of three Kramers degenerate bands, and at the degeneracy point there is also a parabolic hole like band in addition to the four component massless Dirac fermion~\cite{Orlita}.

Recent magnetotransport experiments on Bi$_{1-x}$Sb$_x$, with $x$ tuned at the band inversion transition~\cite{kim1}, and ZrTe$_5$~\cite{kharzeev} have identified a large negative LMR. At very low fields the LMR is positive, which eventually becomes negative with increasing magnetic field strength. In contrast, when $\mathbf{E}$ and $\mathbf{B}$ are orthogonal, these materials show strictly positive transverse magnetoresistance (TMR). The positive LMR for weak magnetic fields is associated with the weak antilocalization effects due to the presence of strong spin orbit coupling in these materials~\cite{hikami}. The analysis of negative LMR data leads to the existence of a large positive LMC varying as $B^2$. In these experiments, the concomitant observations of negative LMR and positive TMR have been claimed to be the experimental verification of the axial anomaly of a WSM~\cite{kim1} (for Bi$_{1-x}$Sb$_x$) or a DSM~\cite{kharzeev} (for ZrTe$_5$). Independent of whether the axial anomaly is causing the negative LMR or not, the very observation of a large LMR does not fit into the theme of classical theory of magnetotransport. A classical theory predicts zero LMR, due to the absence of a Lorentz force when $\mathbf{E}$ and $\mathbf{B}$ are parallel~\cite{Pippard}, and the experiments are clearly pointing toward the existence of a subtle quantum mechanical effect. It is also important to note that the negative LMR is not caused by the field induced suppression of the scattering from the magnetic impurities. Otherwise, even the TMR will decrease as a function of the magnetic field strength, which is not observed in these experiments. The existence of negative LMR, without the notion of axial anomaly was first considered by Adam and Argyres for a three dimensional electron gas sixty years ago~\cite{Argyres}.

However, our particular interest in the potential effects of the axial anomaly on the LMR has been stimulated by the recent magnetotransport measurements on a quasi-two dimensional metal PdCoO$_2$~\cite{balicas}. The Fermi surface of this material is a corrugated cylinder, composed of a single sheet. The \emph{ab initio} calculations~\cite{band3} and the ARPES measurements~\cite{arpes} are in good agreement, and do not show any evidence for the existence of an underlying three dimensional Dirac like band structure. Nevertheless, a large negative LMR and a large positive TMR have been concomitantly observed, when the field is applied along the crystallographic $c$ axis or the interlayer direction. In contrast to the two Dirac materials discussed above (i.e., Bi$_{1-x}$Sb$_x$ and ZrTe$_5$), this delafossite compound demonstrates a negative LMR for weak magnetic fields. The negative LMR becomes more pronounced at low temperatures. Its size and the existence over a wide range of temperatures ($T \leq 200 K$) and magnetic field strengths ($B \leq 30 T$) can not be reconciled with small weak localization corrections, which usually occur at weak fields and low temperatures~\cite{gerd}. In addition, if weak localization is the underlying mechanism, then it will also lead to a negative TMR, in contradiction with the experimental observations. The field dependence of the negative LMR over the wide temperature range  can be fitted with a scaling function of $\omega_c \tau$ according to the Kohler rule~\cite{Pippard}. Here, $\omega_c$ is the cyclotron frequency and $\tau$ is the temperature dependent transport lifetime in the absence of the magnetic field. The data does not suggest a simple $B^2$ behavior of the LMC. These observations have led the authors to propose the axial anomaly as the driving mechanism behind the observed negative LMR. In this context, the axial anomaly has been argued to be a ubiquitous feature of any three dimensional metal in the presence of a quantizing magnetic field, which arises from the quasi one dimensional conduction channels caused by the partially filled Landau levels (LLs) dispersing along the applied magnetic field. This qualitative picture has been further supported by the angle dependent magnetoresistance oscillation (AMRO) measurements. When the magnetic field is tilted with respect to the $c$ axis at certain special angles, known as Yamaji angles~\cite{Yamaji,Kurihara}, the LLs become non-dispersive and the material displays an extremely large and strictly positive interlayer magnetoresistance. As the current is measured along the $c$ axis, which is now at an angle with $\mathbf{B}$, the measured magnetoresistance is a combination of LMR and TMR. When the $B$ field is slightly tilted away from the Yamaji angle, the LLs reacquire their dispersion and a negative component of MR emerges. More strikingly, the negative component overwhelms the existing large and positive component with increasing field strength. This gives strong support to the idea that the axial anomaly arising in the presence of the one-dimensional Fermi points is responsible for the observed negative LMR.

Motivated by these intriguing experimental results, here we develop a comprehensive theory of longitudinal magnetotransport in the strong field limit, $\omega_c \tau \gg 1$. Our main achievements are as follows:
\begin{enumerate}
\item We first establish how the dispersing LLs for a generic metal can lead to the axial anomaly as described by Eq.~(\ref{ABJ2}). One may be surprised by the fact that we are not placing any stringent constraint on the underlying band structure, \emph{i.e., the system does not need to be a WSM or a DSM by any means!} This universality is brought out by the quantizing magnetic field, which is not a weak perturbation. Irrespective of the underlying band structure, the spectrum is bunched into highly degenerate LLs dispersing along $\mathbf{B}$. We show how this field induced dimensional reduction (for the purpose of longitudinal magnetotransport) or quasi-one dimensionality is ultimately responsible for the emergence of the axial anomaly. \\

\item Through an explicit calculation we show how the axial anomaly as a quantum mechanical phenomenon generically leads to LMR, in sharp contrast with the predictions of the classical theory of magnetotransport. We employ a Boltzmann equation approach to calculate the LMC in the presence of short range scattering due to neutral impurities and point defects as well as long range scattering due to ionic impurities. In the quantum limit ($B>B_0$), when only the lowest Landau level (LLL) is partially filled, the ionic impurities make a positive contribution to the LMC $\propto B^2$ for any metal. In contrast, the short range impurities can cause some nonmonotonic behavior of the LMC up to a threshold $B_t$ and an eventual decrease of LMC in the asymptotic limit $B \to \infty$ for a conventional metal. The decrease of LMC for a Gaussian impurity and a point defect vary as $B^{-1}$ and $B^{-2}$ respectively. Therefore, in the presence of ionic impurities alone, the axial anomaly leads to a large unsaturated negative LMR, and a LMC $\propto B^2$ in the quantum limit. This invalidates a common assumption in the literature that the axial anomaly always leads to a $B$-linear positive LMC in the quantum limit.~\cite{aji,son,kim1,ong,shovkovy} Only for a metal with relativistic dispersion, such as WSM and DSM, neutral Gaussian impurities can cause a $B$-linear LMC. In the presence of both ionic and short range scattering, the contributions to the LMR add up according to the the Mathiessen's rule. Consequently, after entering the quantum limit ($B>B_0$), the LMR for a conventional metal with nonrelativistic dispersion first becomes negative and passes through a minimum at a nonuniversal field strength $B=B^\ast$, before its eventual rise to become positive, as illustrated generically in Fig.~\ref{LMR}. \\

\begin{figure}[htb]
\includegraphics[width=8cm,height=6cm]{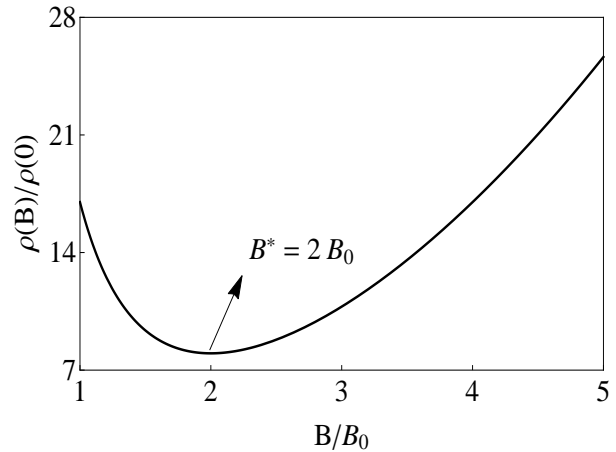}
\caption{An illustration of the LMR due to the combined effects of neutral and ionic impurities [Eq.~(\ref{netLMR})]. The quantum limit is achieved at a magnetic field strength $B=B_0$. The initial decrease of LMR for $B>B_0$ is governed by ionic scattering and attains its minimum value at a nonuniversal field strength $B^\ast$ [Eq.~(\ref{B*})]. For $B>B^\ast$, the sharp increase of LMR is controlled by the neutral impurities. The choice of $B^\ast=2B_0$ has been made only for an illustrative purpose. The actual value of the crossover field $B^\ast$ depends on many nonuniversal sample-dependent details such as the comparative amounts of ionic long-range and neutral short-range disorder, the carrier density, etc., and $B^\ast$ could, in principle, be large or small, thus suppressing the positive or negative LMR regimes in experiments.}
\label{LMR}
\end{figure}

\item We explicitly discuss the Landau level (LL) structures of various physical systems such as a three dimensional electron gas, a quasi two dimensional metal, a WSM, and a DSM with an even as well as an odd number of flavors. For each of these systems we show the validity of our general results regarding the LMC in the quantum limit.\\

\item For layered materials with a corrugated Fermi surface, we provide a full quantum mechanical treatment for establishing the existence of the Yamaji angle in a tilted magnetic field with respect to the interlayer direction. We explicitly show that for a very small deviation from the Yamaji angle, a single LL remains partially filled, and the system effectively realizes a situation similar to the one in the quantum limit. This happens only for the layered materials with a large carrier density, such that the energy scale associated with the areal density is larger than the interlayer hopping strength. Note that only for such a high carrier density, the underlying Fermi surface can be a corrugated cylinder, as found in PdCoO$_2$ and PtCoO$_2$. Other layered materials such as graphite~\cite{behnia1} and $\alpha$-(BEDT-TTF)$_2$I$_3$ can also exhibit a negative LMR in certain range of the field strength. However, these are compensated semimetals with very low carrier density and a set of closed electron and hole pockets. Our theory provides a simple explanation why the Yamaji angle and its associated phenomena will be absent in such materials with low carrier density.\\

\item We discuss the experiments on Bi$_{1-x}$Sb$_x$~\cite{kim1}, and ZrTe$_5$~\cite{kharzeev} in light of our theory. We show that the experimental observation of positive LMC $\propto B^2$ in these semiconducting materials are consistent with our theory in the quantum limit, with ionic scattering being the dominant relaxation mechanism, as is generically expected in semiconductors.
    We also argue why the observation of the positive LMC in ZrTe$_5$~\cite{kharzeev}, can not by itself, be construed as evidence for the existence of the chiral magnetic current.\\
\end{enumerate}

The manuscript is organized as follows. In Sec.~\ref{1danomaly} we demonstrate the relation between the axial anomaly as described by Eq.~(\ref{ABJ1}) and the electrical conductivity of a one dimensional system. We also show why the notion of the axial anomaly can be applied even for a bounded dispersion relation. In Sec.~\ref{sec:3danomaly}, we explain how the one dimensional anomaly of a dispersive LL is responsible for giving rise to the axial anomaly (see Eq.~(\ref{ABJ2})) for a generic three dimensional system in parallel electric and magnetic fields. We demonstrate how the ABJ anomaly gives rise to the LMC and the LMR in contrast to the predictions of the classical theory of magnetotransport. We again ensure that our conclusions are valid for bounded dispersion relations. In Sec.~\ref{sec:LMCQL}, we calculate the LMC in the quantum limit, using a Boltzmann equation in the presence of short range and long range scattering due to neutral and ionic impurities respectively. In the Sec.~\ref{subsec:electrongas}, we consider the LMC in the quantum limit for a three dimensional electron gas, as found in many semiconductors. The discussion of the LMC for quasi-two dimensional or layered metals are presented in Sec.~\ref{sec:QL}. We consider many aspects of the magnetotransport experiments on PdCoO$_2$ in Subsec.~\ref{delafossite}. The agreement between our theory and the experimentally observed LMR in $\alpha$-(BEDT-TTF)$_2$I$_3$ is discussed in Subsec.~\ref{organic}. The LMC of a WSM and a DSM with an even number of flavors are considered in Secs.~\ref{subsec:Weyl} and ~\ref{subsec:2Dirac} respectively. A brief discussion of the magnetotransport experiments on Cd$_3$As$_2$ can be found in Sec.~\ref{subsec:2Dirac}.  Sec.~\ref{subsec:1Dirac} is devoted to the discussion of a DSM with an odd number of flavors. In particular, we discuss recent magnetotransport measurements on Bi$_{1-x}$Sb$_x$ and ZrTe$_5$ in Subsec.~\ref{bisb} and Subsec.~\ref{subsec:CME} respectively. In Subsec.~\ref{subsec:CME} we show why the observation of a positive LMC varying as $B^2$ does not necessarily imply the existence of a chiral magnetic current.
Our main findings are summarized in Sec.~\ref{Conclusions}. An explicit derivation of the transport lifetime using the Boltzmann equation is relegated to the Appendix~\ref{boltzmannLMC}.

\section{Axial anomaly of one dimensional metals} \label{1danomaly}
Quite amazingly, the axial anomaly described by Eq.~(\ref{ABJ1}) can naturally arise in any one dimensional electron gas due to the existence of Fermi points and linearly dispersing low energy excitations in their vicinity. This one dimensional form of the axial anomaly will play a crucial role in the subsequent development of the theory of LMR in three dimensions in the presence of an external magnetic field. For this reason, we will first consider the implications of the axial anomaly for the electrical conductivity of a one dimensional system of spinless, charged fermions with dispersion $\epsilon(k)$.

The low energy physics is governed by the linearly dispersing right and left movers in the vicinity of the two Fermi points located at $k=\pm k_F= \pm \pi n$, where $n$ is the linear density of fermions. The effective action for the linearized theory is given by
\begin{equation}
S=\int dx dt \; \psi^\dagger \left[i\hbar \partial_t + i\hbar v_F \tau_3 \partial_x \right]\psi.
\end{equation} Here $\tau_3=\mathrm{diag}(1,-1)$ is a Pauli matrix, and $\psi^T=(R,L)$ is a two component Grassmann spinor, with $R$ and $L$ respectively describing the right and the left movers, which are one component Weyl fermions. The Fermi velocity is defined as $\hbar v_F=\partial_k \epsilon (k)|_{k=k_F}$. The linearized theory is invariant under two global gauge transformations $\psi \to e^{i\alpha} \psi$ and $\psi \to e^{i\alpha_{ax} \tau_3} \psi$, which implies separate number conservation laws for the right and left movers. The total or electrical charge and current density operators are defined as
\begin{eqnarray}
\rho(x,t)=-e \psi^\dagger \psi, \: j(x,t)=-ev_F \psi^\dagger \tau_3 \psi.\nonumber \\
\end{eqnarray} In contrast, the axial charge and current density operators are defined as
\begin{eqnarray}
\rho_{ax}(x,t)=-\frac{j(x,t)}{ev_F}, \: \: j_{ax}=\frac{v_F \rho(x,t)}{e}.\label{eq5}
\end{eqnarray} These relations between the total and the axial current operators are special properties of a two component Dirac fermion in one dimension. In the presence of an electric field $E$, the electrical charge and current still satisfy the continuity equation
\begin{equation}
\partial_t \rho + \partial_x j=0.
\end{equation}
In contrast, the axial current conservation law is violated according to Eq.~(\ref{ABJ1}), with $N_f=1$. Notice that the spin degree of freedom can be taken into account by setting $N_f=2$. In realistic solid state systems, $N_f$ could be a larger integer because of valley degeneracy and band structure effects. We also note that the axial anomaly equation serves as the basis for the bosonization procedure.

The physical consequence of the axial anomaly can be better understood by considering a constant external electric field. Such a field does not induce an electrical charge density $\rho$, which naturally sets $j_{ax}=0$. After integrating both sides of Eq. (\ref{ABJ1}) with respect to time, we find a linear growth of the axial charge and the electrical current with time, described by
\begin{equation}
\rho_{ax}=\frac{j}{ev_F}=\frac{eEt}{\pi \hbar},
\end{equation} for $N_f=1$. The linear time dependence of $j$ corresponds to a uniform acceleration of the center of mass. However, in the presence of impurity scattering or a periodic potential due to the underlying lattice, this uniform acceleration can not be sustained for an indefinitely long time. For a generic solid state system, the transport lifetime $\tau_{tr}$ due to the impurity scattering is much shorter than the Bloch oscillation period $T=\hbar/(eEa)$ due to the underlying periodic potential, where $a$ is the lattice spacing. Therefore, the anomaly equation will be pertinent for a time scale $t \leq \tau_{tr} \ll T$. Since an axial charge build up is occurring between the two Fermi points, it is natural to anticipate the relaxation rate to be related to the backscattering rate. Based on this intuition, we can use the following phenomenological equation to describe the relaxation of the axial charge density
\begin{equation}
\partial_t \rho_{ax}=\frac{eE}{\pi \hbar}-\frac{\rho_{ax}}{\tau_{b}}. \label{1daxialrelaxation}
\end{equation} In the steady state, this gives rise to a remarkable relation between the axial charge and the electrical conductivity
\begin{equation}
\sigma=\frac{ev_F\rho_{ax}}{E}=\frac{e^2v_F\tau_{b}}{\pi \hbar},\label{1daxialconductivity}
\end{equation} where $\tau^{-1}_b$ is the backscattering rate. We will be using this relation for the subsequent development of the axial anomaly for three dimensional metals and their magnetoconductivity. Therefore, we will first confirm the validity of this phenomenological description, by explicitly using the solution of the Boltzmann equation for the microscopic model, without linearizing the dispersion relation. This will also provide us with a concrete formula for the relaxation rate.

Within the relaxation time approximation, the steady state current for the microscopic model is given by
\begin{eqnarray}
j&=&\sigma E=\frac{e^2\tau_{tr}E}{\pi}\int \frac{dk}{2\pi} \; v^2(k) \; \delta(\epsilon_F-\epsilon(k))\nonumber \\
&=&\frac{e^2 v_F \tau_{tr}}{\pi \hbar}E,\label{1dboltzmannconductivity}
\end{eqnarray} where $v(k)=\partial_k \epsilon(k)/\hbar$ is the group velocity. Notice that the same expression for the conductivity is predicted by the phenomenological description of the axial charge relaxation if we identify $\tau_{tr}=\tau_b$. Further confirmation of this comes from the solution of the linearized Boltzmann equation
\begin{eqnarray}
\frac{1}{\tau_{tr}}=\int \frac{dq}{2\pi} W(k,k^\prime) \left[1-\frac{v(k+q)}{v(k)}\right].\label{1dlinearizedboltzmann}
\end{eqnarray} The collision processes are captured by
\begin{eqnarray}
W(k,k+q)=\frac{2\pi}{\hbar}n_i|U_{1d}(q)|^2\delta(\epsilon(k)-\epsilon(k+q)),
\end{eqnarray} where $n_i$ is the linear density of the impurities and $U_{1d}(q)$ is the Fourier transform of the one dimensional impurity potential. The linearized Boltzmann operator is annihilated by the forward scattering processes at $q=0$, and the transport lifetime is entirely determined by the backscattering at $q=-2k$. This now provides a proper justification for identifying the transport lifetime with the backscattering lifetime, and we obtain the following expression
\begin{eqnarray}
\tau_{tr}=\tau_{b}=\frac{\hbar^2 v_F}{2n_i |U_{1d}(2k_F)|^2}.\label{1dbackscattering}
\end{eqnarray}
Therefore, we can use the phenomenological description of Eq.~(\ref{1daxialconductivity}), together with Eq.~(\ref{1dbackscattering}), to describe the anomaly induced electrical transport in one dimension. Now we are in a position to address the relevance of ABJ anomaly for a three dimensional system in the presence of parallel electric and magnetic fields.

\begin{figure}[htb]
\includegraphics[width=8cm,height=6cm]{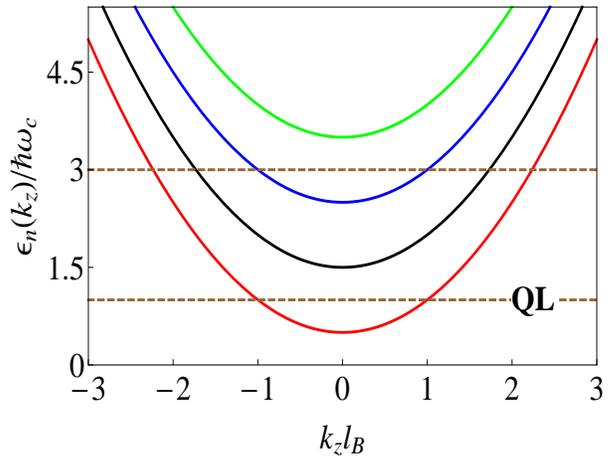}
\caption{(Color online) The dispersing LLs of a three dimensional electron gas, where $\omega_c=eB/m$ is the cyclotron frequency and $l_B=\sqrt{\hbar/eB}$ is the magnetic length. The LLs with $N=0,1,2$ and $3$ are respectively represented with the colors red, black, blue and green. The dashed brown lines correspond to two possible locations of the Fermi energy. When the Fermi energy lies entirely within the $N=0$ LLL, we mark it by QL to illustrate that we are in the quantum limit. In the quantum limit only the partially occupied LLL produces a set of Fermi points. In contrast, away from the quantum limit, several partially filled LLs can produce Fermi points. Each partially occupied LL gives rise to the axial anomaly and participates in the longitudinal magnetotransport.}
\label{electrongasLandaulevels}
\end{figure}

\section{Axial anomaly of three dimensional metals} \label{sec:3danomaly}
We will first consider a generic three dimensional metal in the presence of a strong magnetic field $\mathbf{B}=(0,0,B)$. For simplicity we will assume the existence of a closed Fermi surface contour in the plane perpendicular to $\mathbf{B}$. In this case, the cyclotron motion is quantized into discrete LLs, which can only disperse along the direction of the magnetic field. In the Landau gauge, $\mathbf{A}=(0,Bx,0)$, we can describe the dispersion relation of the LLs according to
\begin{equation}
\epsilon_N(k_y,k_z)=\epsilon_N(k_z),
\end{equation} which at least has the degeneracy $eB/h$, with a magnetic length $l_B=\sqrt{\hbar/eB}$. There may be additional degeneracy of the LLs, arising from the presence of multiple atoms in a unit cell, which we will denote by $g_N$ for each level. A partially filled LL at least crosses the Fermi energy at a set of Fermi points $\pm k_{F,N}(B)$, and acts as a highly degenerate quasi one dimensional conduction channel. For example, we show a few LLs of a three dimensional electron gas in Fig.~\ref{electrongasLandaulevels}. When an external electric field is applied parallel to the magnetic field, there is no Lorentz force in this direction, and classically one does not expect any LMC or LMR. Only the partially filled LLs possessing Fermi points can participate in the longitudinal magnetotransport, and following the discussion in the previous section, we anticipate an important connection between the one dimensional form of the axial anomaly and the LMC.

Now consider the quantum limit, achieved for a very strong magnetic field strength $B>B_0$, such that all the carriers are confined to the LLL with $N=0$. The field scale $B_0$ is determined by the density of carriers and the parameters of the underlying microscopic model. In the absence of
 impurity scattering, the axial anomaly formula in the quantum limit is obtained by multiplying the right hand side of Eq.~(\ref{ABJ1}) by $eB/h$ and identifying $N_f$ as the orbital or spin degeneracy $g_0$ in the LLL, which leads to
\begin{equation}
\partial_{t} \rho_{ax,0}=g_0  \frac{e^2 E B}{2\pi^2 \hbar^2}.\label{ABJ0}
\end{equation}
Notice that we have obtained the ABJ anomaly formula of Eq.~(\ref{ABJ2}) without referring to any underlying relativistic band structure. Therefore, the emergence of the axial anomaly is a universal feature of all three dimensional metals placed in parallel electric and magnetic fields. In the presence of impurity scattering, the axial charge and the magnetoconductivity in the quantum limit can be determined by following Eq.~(\ref{1daxialconductivity}). They are given by
\begin{eqnarray}
&&\rho_{ax,0}=g_0\frac{e^2 E B \tau_{b,0}(B)}{2\pi^2 \hbar^2},\label{axialb0}\\
&&\sigma(B)=g_0\frac{ e^2 v_{F,0}(B)\tau_{b,0}(B)}{2\pi^2 \hbar l^2_B} , \label{sigmab0}
\end{eqnarray} where $\tau^{-1}_{b,0}$ is the backscattering rate between the Fermi points and $v_{F,0}(B)$ is the Fermi velocity
 in the LLL. The backscattering rate, which depends on the nature of disorder, has to be determined from the Boltzmann equation for longitudinal magnetotransport in the LLL. The evaluation of $\tau_{b,0}$ is shown in the following section. The formula for LMC in Eq.~(\ref{sigmab0}) is one of the main results of this
 manuscript and is applicable to any weakly correlated generic metal in the quantum limit, independent of the explicit form of the band dispersion. This is the generalization of Nielsen-Ninomiya's formula~\cite{nielsen} for axial anomaly induced LMC of Weyl fermions to a generic three dimensional meta. As long as $v_{F,0}(B)\tau_b(B)/l^2_B$ is not a constant, the axial anomaly can give rise to LMC and LMR, in contrast to the anticipations of the classical theory of magnetotransport. The issue of positive or negative LMR is determined by the explicit dependence of the relaxation rate on $B$, depending explicitly on the nature of the underlying disorder, which we will consider in the next section.

 It has been known for a long time that the dimensional reduction caused by the strong magnetic field in the quantum limit, can also lead to a charge density wave (CDW) instability of the electronic system at very low temperatures~\cite{fukuyama,Sankar1,bryant}.
 Such instabilities generically tend to gap out the Fermi points, and will lead to an insulating behavior of the LMR for a clean, interacting system. However, it is important to note that an experimental detection of such a broken symmetry phase is quite challenging and signatures of certain instabilities have only been reported in graphite and bismuth. Our theory is based on the existence of the Fermi points in the absence of any such instability. This assumption can be justified in the following way. The disorder potential couples to CDW order as a random field, and it generally plays a detrimental role for the existence of a CDW as a true long range order. If we consider a continuum model of the electron gas, a charge density wave breaks the continuous translational symmetry, and the Imry Ma argument forbids the existence of a continuous symmetry breaking below four dimensions in the presence of a random field. In contrast, for a lattice model of electrons, a CDW only breaks a discrete translational symmetry, and  the Imry Ma argument does not prohibit its existence as a long range order in three dimensions. However, disorder can cause a severe suppression of the transition temperature~\cite{Sankar1}, while making it effectively impossible to detect such ordering in most materials. In any case, all our results would remain valid at temperatures higher than the transition temperature of such a CDW instability (in the unlikely scenario that an instability does occur), and we ignore the possibility of a CDW instability in the current work.

For weaker magnetic field strengths below the quantum limit $B<B_0$, there are multiple partially occupied LLs. Each of these levels will have their own ABJ anomaly defined by the equation
\begin{equation}
\partial_{t} \rho_{ax,N}=g_N  \frac{e^2 E B}{2\pi^2 \hbar^2},\label{ABJN}
\end{equation}
where $g_N$ is the orbital and/or spin degeneracy in the $N$th LL. Thus each LL contributes to the LMC through its transport lifetime $\tau_{tr,N}(B)$ and Fermi velocity $v_{F,N}(B)$. This leads to a very general formula for the LMC
\begin{equation}
\sigma(B)=\sum^{\prime}_N g_N \frac{e^2 v_{F,N}(B)\tau_{tr,N}(B)}{2\pi^2 \hbar l^2_B},\label{sigmabN}
\end{equation}
where the prime denotes a sum carried over all the partially filled LLs. The transport lifetimes of various LLs follow a set of coupled linear algebraic equations, which account for both intra and inter LL scattering processes. This will be derived in the Appendix~\ref{boltzmannLMC}. In the Boltzmann theory, the broadening of the LLs is not taken into account. Therefore, Eq.~(\ref{sigmabN}) will lead to undamped Shubnikov de Hass oscillations of the LMR. The inclusion of the broadening at a phenomenological level will have two important consequences: (i) a Dingle factor for each level, which will suppress the amplitude of the oscillations, and (ii) a quantitative modification of the transport lifetimes due to the forward scattering process. However, the backscattering processes with large momentum transfer still make the dominant contribution to the transport lifetime. However, these quantitative details do not alter the fact that \emph{the axial anomaly gives rise to the very existence of a finite LMR}.

We can further demonstrate that Eq.~(\ref{sigmabN}) is not an artefact of a linearized theory, and is valid for an arbitrary LL dispersion. In the presence of parallel electric and magnetic fields, the current carried by the quasiparticles residing in the partially filled LLs are given by
\begin{eqnarray}
j_z=-\frac{e}{2\pi l^2_B}\int \frac{dk_z}{2\pi} \sum^{\prime}_{N} g_N v_{N}(k_z)f_N(k_z).\label{jz}
\end{eqnarray} In the relaxation rate approximation, the deviation from equilibrium is described by
\begin{equation}
f_N(k_z)=f^0_N(k_z)-eEv_N(k_z)\tau_{tr,N}(k_z)\frac{\partial f^0}{\partial \epsilon}|_{\epsilon=\epsilon_N(k_z)}.
\end{equation} After substituting this in Eq.~(\ref{jz}), we obtain
\begin{eqnarray}
j_z=\frac{e^2E}{2\pi^2\hbar l^2_B}\sum^{\prime}_{N} g_N v_{F,N}\tau_{tr,N}(B),
\end{eqnarray}which indeed leads to the most general expression for the LMC announced in Eq.~(\ref{sigmabN}).

It is possible to reach the quantum limit without using very large magnetic fields in certain narrow-gap and gapless semiconductors with very low carrier densities. It is also possible to attain the condition of a single partially filled LL in a quasi-two dimensional metal with high carrier density. This condition can be realized either by making the cyclotron energy bigger than the interlayer hopping strength or adjusting the angle between the applied magnetic and electric fields in the proximity of the ``Yamaji angles''.  We discuss these aspects in detail in the Sec.~\ref{sec:QL}. In the next section we focus on determining the transport lifetime and the LMC in the quantum limit.

\section{Longitudinal magnetoconductivity in the quantum limit} \label{sec:LMCQL}
Computing the transport lifetime from the Boltzmann equation, within the relaxation time approximation for a generic dispersion $\epsilon_{N=0}(k_z)$ in the quantum limit, we arrive at
\begin{equation}
\tau_{b,0}(B)=\frac{\hbar^2 v_{F,0}(B)}{2 n_i l^2_B |U_{eff}(2k_{F,0}(B))|^2},
\end{equation}where $n_i$ is the impurity density per unit volume and $U_{eff}(q_z)$ is an effective one dimensional scattering potential. This scattering potential has to be obtained by integrating the matrix elements of the three dimensional disorder potential over the in-plane components of the momentum, and contains information regarding the wavefunction of the LLL. Remarkably, the result of the Boltzmann calculation in the presence of the magnetic field agrees with the one dimensional result of Eq. (\ref{1dbackscattering}), where the one dimensional potential $U_{1d}$ is replaced by $U_{eff}$. The expression for the effective potential is given by
\begin{equation}
|U_{eff}(q_z)|^2=\int \frac{d^2q_\perp}{(2\pi)^2l^2_B}|U(q_\perp,q_z)|^2  \exp \left (-\frac{q^2_\perp l^2_B}{2}\right ),\label{U0qz}
\end{equation}where $U(q_\perp,q_z)$ is the Fourier transform of the actual three dimensional real-space disorder potential. We now consider two types of disorder potentials: (i) short range Gaussian impurity scattering specified by $U^s(q_\perp,q_z)=U_0 \exp[-a^2(q^2_\perp+q^2_z)/2]$, where $a$ is the range of the impurity potential, and (ii) long range ionic impurity scattering characterized by the screened Coulomb potential $U^c(q_\perp,q_z)=(4\pi e^2/\kappa) [q^2_\perp+q^2_z+q^2_{TF}]^{-1}$, where $\kappa$ is the background lattice dielectric constant of the system and $1/q_{TF}$ is the Thomas-Fermi screening length.  We note that in the limit $a\rightarrow 0$ the Gaussian potential reduces to an onsite Dirac delta function potential $U_0\delta({\bf r})$ representative of the zero-range disorder associated with neutral point defects.

For the Gaussian potential, we find the following effective one dimensional potential and the transport lifetime
\begin{eqnarray}
&&U^s_{eff}(q_z)=\frac{U_0}{\sqrt{2\pi l^2_B(l^2_B+2a^2)}} \exp \left(-\frac{q^2_z a^2}{2}\right), \label{ugs}\\
&&\tau^s_{b,0}(B)=\frac{\pi \hbar^2}{n^s_i U^2_0}v_{F,0}(B)(l^2_B+2a^2)\exp[4k^2_{F,0}(B)a^2],\nonumber \\
\end{eqnarray} where $n^s_i$ is the density of neutral Gaussian impurities. Therefore, the contribution of the Gaussian short-range impurity scattering to the LMC in the quantum limit is given by
\begin{eqnarray}
\sigma^s(B)=\frac{e^2\hbar}{2\pi n^s_iU^2_0}v^2_{F,0}(B)\left(1+\frac{2a^2}{l^2_B}\right)\exp[4k^2_{F,0}(B)a^2]. \label{sigmab0short}\nonumber \\
\end{eqnarray} Depending on the underlying dispersion relation, $v_{F,0}$ can be a periodic function of $k_{F,0}(B)$. This in turn can cause a nonmonotonic dependence of $\sigma(B)$ on the magnetic field strength. A detailed discussion of the nonmonotonic variation of the LMC due to zero range impurity scattering or point defects $(a \rightarrow 0)$ will be provided in the following sections, when we consider different physical systems. However, for a system with nonrelativistic dispersion and very large magnetic field strength, the asymptotic behavior of $\sigma^s(B)$ is given by $$\sigma^s(B) \sim l^2_B(l^2_B+2a^2).$$ Therefore, the short range scattering due to neutral impurities always lead to positive LMR for a conventional metal in the very large magnetic field limit $B \to \infty$. In contrast, the finite range Gaussian impurities can lead to a $B$-linear positive LMC for systems with underlying relativistic dispersion, which will be discussed in the following sections.

For long-range (i.e. Coulombic) ionic impurity scattering, the effective one dimensional potential is given by
\begin{eqnarray}
U^c_{eff}(q_z)=\frac{\sqrt{2\pi}e^2}{\kappa}\sqrt{\Gamma [-1,(q^2_z+q^2_{TF,0}(B))l^2_B/2 ]} \times \nonumber \\ \exp [(q^2_z  +q^2_{TF,0}(B))l^2_B/4],\label{ucs}
\end{eqnarray}where $\Gamma(-1,x)$ is the upper incomplete Gamma function, and $q_{TF,0}^{-1}$ is the Thomas Fermi screening length in the quantum limit. By using this effective potential we find the following contribution due to the ionic impurities to the LMC
\begin{eqnarray}
\sigma^c(B)=\frac{\hbar \kappa^2  v^2_{F,0}(B)}{8\pi^3 e^2 n^c_i  l^4_B}\frac{\exp \left[-\frac{l^2_B}{2} \{4k^2_{F,0}(B) +q^2_{TF,0}(B) \} \right]}{ \Gamma \left[-1,\frac{l^2_B}{2} \{4k^2_{F,0}(B)+q^2_{TF,0}(B) \} \right]}, \label{ionicmagnetoconductivity}\nonumber \\
\end{eqnarray} where $n^c_i$ is the density of the charged impurities. The analysis of the asymptotic behavior of $\sigma(B)$ can be facilitated by using the following relation
\begin{equation}
e^x \Gamma(-1,x)=\sum_{n=0}^{\infty}(-1)^n \Gamma(n+2)x^{-(n+2)}.
\end{equation} In the quantum limit, the enhanced degeneracy of the LLL leads to very strong screening of the ionic impurities, which is captured through the following field dependent Thomas Fermi momentum
\begin{equation}
q_{TF,0}(B)=\frac{e}{\sqrt{2\pi^2 \hbar v_{F,0}(B)l^{2}_B}}.\label{qTF}
\end{equation} Since $q_{TF,0}$ and $k_{F,0}$ are respectively increasing and decreasing functions of $B$, the asymptotic expansion is dominated by $q_{TF,0}$. This leads to a strong positive LMC increasing with field strength as $\sim B^2$. This is the generic high-field, extreme quantum limit behavior of charged disorder induced LMC in three-dimensional metals.

However, this asymptotic behavior can be understood on a more physical ground. Since the screening length in the quantum limit decreases with the increasing magnetic field strength, the effective one dimensional scattering potential also decreases as $U^{c}_{eff} \sim q^{-2}_{TF,0}l^{-1}_B$. Consequently, the relaxation time $\tau^{c}_{b,0} \propto v_{F,0} (U^{c}_{eff})^{-2}$ increases with $B$. As the conductivity in the asymptotic limit $B \to \infty$ behaves as $\sigma^c(B) \sim v^2_{F,0}q^4_{TF,0}$, using Eq.~(\ref{qTF})
we arrive at one of our key results
$$\sigma^c(B) \sim l^{-4}_{B}\sim B^2,$$
which is valid for an arbitrary dispersion relation of the LLL. When ionic scattering is the only resistive relaxation mechanism, there is a large negative LMR in the quantum limit, and the positive LMC $\sigma^c(B) \propto B^2$ describes a universal result.
\begin{widetext}
\begin{figure*}[tb]
\centering
\subfigure[]{
\includegraphics[scale=0.75]{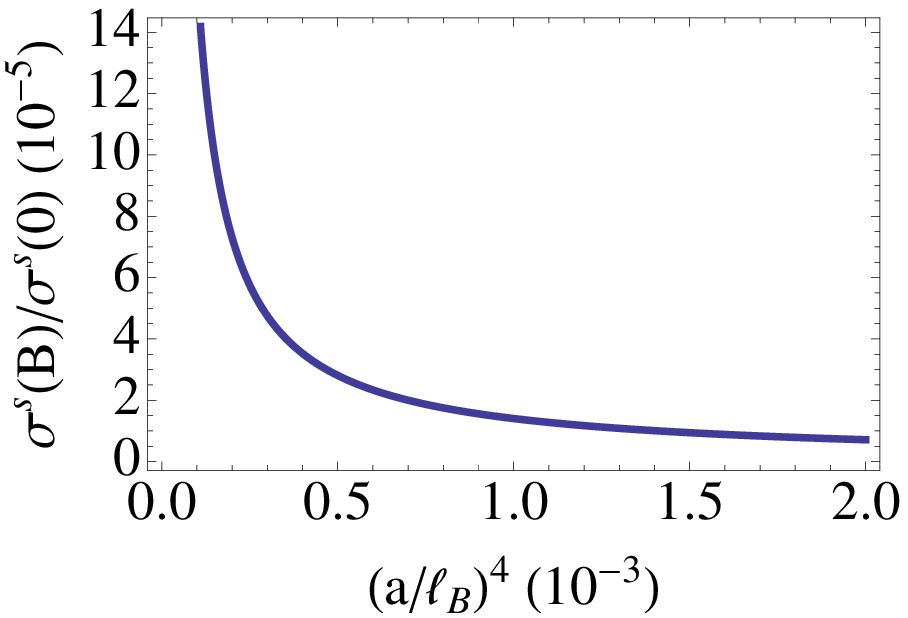}
\label{fig:GS}
}
\subfigure[]{
\includegraphics[scale=0.75]{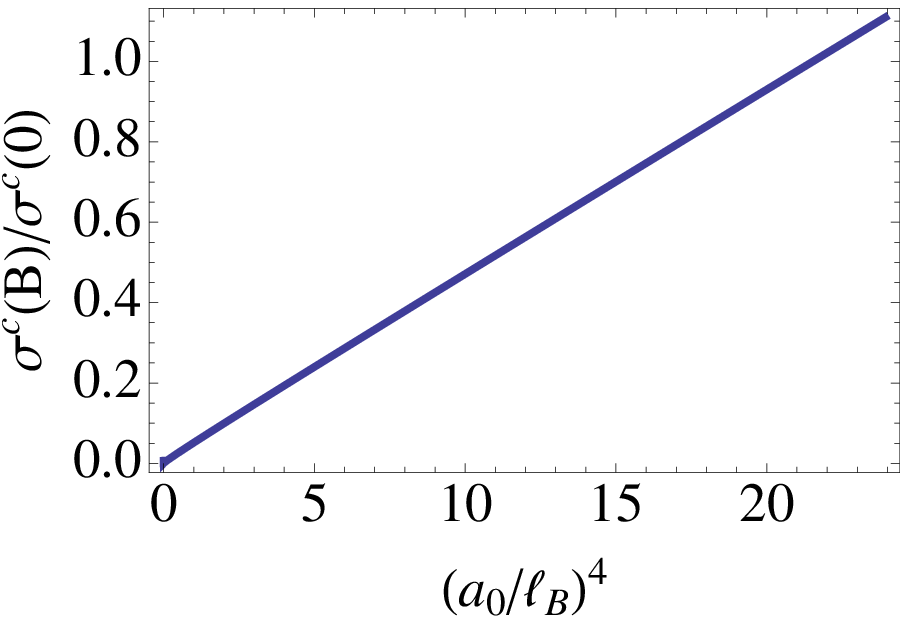}
\label{fig:CS}
}
\label{fig:LMCelectrongas}
\caption[]{LMC for an electron gas in semiconductors [Eq.~(\ref{eqn:quad_en})] in the quantum limit. (a) Gaussian potential scattering: LMC with a fixed carrier density of $n=6.75\times10^{26}m^{-3}$ as a function of $(a/l_B)^4$, where $a$ is the range of the scattering potential, and $\sigma^s(0)$ is the conductivity in the absence of a magnetic field due to short range scattering (see the Appendix). We find $\sigma^s(B)/\sigma^s(0)$ decreases as a function of $B$, which gives rise to a positive LMR. (b) Ionic scattering: LMC with a fixed carrier density of $n=6.75\times10^{26}m^{-3}$ as a function of $(a_0/l_B)^4$, where $a_0$ is the Bohr radius in the vacuum and $\sigma^c(0)$ is the conductivity in the absence of a magnetic field (see the Appendix). We find $\sigma^c(B)/\sigma^c(0) \sim B^2$, which gives rise to a negative LMR.
}
\end{figure*}
\end{widetext}

In the presence of both types (i.e. long-range ionic and short-range neutral) of impurities, the net relaxation rate is given by Matthiessen's rule
\begin{eqnarray}
\frac{1}{\tau_{b,0}(B)}=\frac{1}{\tau^{c}_{b,0}(B)}+\frac{1}{\tau^{s}_{b,0}(B)},
\end{eqnarray} which leads to the net LMR
\begin{equation}
\rho(B)=\rho^c(B)+\rho^s(B).
\end{equation} Here $\rho^c(B)=1/\sigma^c(B)$ and $\rho^s(B)=1/\sigma^s(B)$. When $\sigma^c(B) \ll \sigma^s(B)$, we obtain a negative LMR and a positive LMC $\sigma(B) \propto B^2$. For magnetic field strengths above a nonuniversal threshold $B^\ast$, the contribution from short range impurity dominates, which leads to a positive LMR for a conventional metal. Therefore, short range impurities always become important for an asymptotically large field strength. The value of $B^\ast$ depends on the ratio of the impurity densities ($n^c_i/n^s_i$), and various material parameters such as effective mass and density of carriers. For concreteness, if we consider only zero-range point impurities as the source of short range scattering, we find
\begin{eqnarray}
&& \rho(B)\approx  \frac{\rho(B^\ast)}{2} \left(\frac{B^\ast}{B}\right)^2\left[1+\left(\frac{B}{B^\ast}\right)^4\right], \label{netLMR}\\
&& B^\ast=  \frac{h^2\sqrt{\pi n}}{e\sqrt{U_0\kappa m_{z}}} \left(\frac{n^c_i}{n^s_i}\right)^{1/4},\label{B*}
\end{eqnarray} where $n$ is the density of the carriers and $m_z$ is the effective band mass along the direction of the applied field. Notice that we are defining $B^\ast$ through the condition $\rho^c(B^\ast)=\rho^s(B^\ast)=\rho(B^\ast)/2$. The generic behavior of the LMR described by Eq.~(\ref{netLMR}) is illustrated in Fig.~\ref{LMR}.

In the following sections we establish the applicability of our theory for various experimentally relevant materials. We consider different physical systems according to the varying degree of complexity of the underlying band structure. We begin with the simplest situation of a three dimensional electron gas with parabolic dispersion, followed by the consideration of a simple quasi two dimensional metal. In the absence of the magnetic field, both systems have single sheet of Fermi surface. Later we consider the examples of WSM and DSM, which can possess multiple Fermi surfaces due to the additional orbital degrees of freedom and spin-orbit coupling.

\section{Electron gas in semiconductors} \label{subsec:electrongas}
Usually semiconductors have low carrier density, and for this reason it is easier to achieve the quantum limit even with a moderately strong magnetic field. Within the $\mathbf{k} \cdot \mathbf{p}$ theory, the carriers in many semiconductors can be modeled by a conventional three dimensional electron gas with a parabolic dispersion. The pertinent Hamiltonian in the presence of the magnetic field is (ignoring spin)
\begin{eqnarray}
H_1=\frac{1}{2m}(\mathbf{p}-e\mathbf{A})^2,
\end{eqnarray}where $m$ is the effective band mass. In this case, the LLs have the following dispersion relations
\begin{equation}
\epsilon_N(k_y,k_z)=\left(N+\frac{1}{2}\right)\hbar \omega_c +\frac{\hbar^2 k^2_z}{2m},
\label{eqn:quad_en}
\end{equation} where the cyclotron frequency $\omega_c=eB/m$. First four dispersing LLs and their possible intersections with the Fermi level are illustrated in Fig.~\ref{electrongasLandaulevels}. The quantum limit is achieved when the field strength $B$ exceeds the threshold value
\begin{equation}
B_0=\frac{\hbar}{e} \left(2\pi^4 n^2\right)^{1/3}.
\label{eqn:b0}
\end{equation} The Fermi wavenumber for this problem decreases with $B$ according to
\begin{equation}
k_{F,0}=\frac{mv_{F,0}}{\hbar}=2\pi^2 n l^2_B, \label{kf0}
\end{equation} and the Thomas Fermi screening length is determined by
\begin{equation}
q_{TF,0}^{-1}= \frac{\pi h \sqrt{n}l^2_B}{e\sqrt{m}}\label{parabolicscreening}.
\end{equation}

For short range impurity scattering, we find
\begin{eqnarray}
\sigma(B)=\frac{2e^2\pi^3a^2n^2\hbar^3}{n_iU^2_0m^2}l^2_B\left[1+\frac{l^2_B}{2a^2}\right]\exp(16\pi^4n^2a^2l^4_B), \label{eq:GS}\nonumber \\
\end{eqnarray}
which is a monotonically decreasing function of $B$ (see Fig.~\ref{fig:GS}). For a finite range of the impurity potential, the LMC deceases as $ \sim B^{-1}$. While for zero-range point impurities ($a \to 0$), the LMC decreases as $\sim B^{-2}$. In contrast, the axial anomaly in the presence of long-range ionic impurity scattering leads to a large positive LMC growing as $\sim B^2$ (see Fig.~\ref{fig:CS}). After substituting the expressions for $k_{F,0}$, $v_{F,0}$ and $q_{TF,0}$ in Eq.~(\ref{ionicmagnetoconductivity}), we obtain an explicit form of the magnetoconductivity due to the ionic impurities, which precisely agrees with the formula derived by Adam and Argyres~\cite{Argyres}. After combining the contributions from both types of impurity scattering, we obtain the net LMR given by Eq.~(\ref{netLMR}), where
\begin{equation}
B^\ast=B_0 \frac{1.92 \; h }{\sqrt{U_0\kappa m \; n^{1/3}}} \left(\frac{n^c_i}{n^s_i}\right)^{1/4}.
\end{equation} Therefore, a low density of carriers simultaneously decreases $B_0$ and increases the crossover field strength $B^\ast$, which is suitable for observing the negative LMR over a significant range of the applied field strength. Due to the presence of donors and acceptors in semiconductors, the scattering by ionic impurities is the most important relaxation mechanism at low temperatures. Therefore, semiconducting systems will provide many suitable candidate materials for observing the axial anomaly induced negative LMR in the quantum limit. In addition, semiconductors, because of their low carrier density, can be easily driven into the quantum limit of the LLL occupancy, necessary for the generic axial anomaly being discussed in this work.

An explicit consideration of the Zeeman spin splitting does not change our conclusions. In the absence of spin flip scattering, there is no impurity induced matrix elements between two Zeeman split LLs in the LMC calculation. Therefore, our calculations go through for each spin split LL without any essential modification. We only need to ensure that the total density of carriers is distributed between two spin split levels. In the very strong field limit, eventually one arrives at a fully spin polarized limit, which is exactly described by our calculations for a spinless model. Therefore, our results properly capture the LMC in the asymptotic limit $B \to \infty$.


\begin{widetext}
\begin{figure*}[tb]
\centering
\subfigure[]{
\includegraphics[scale=0.65]{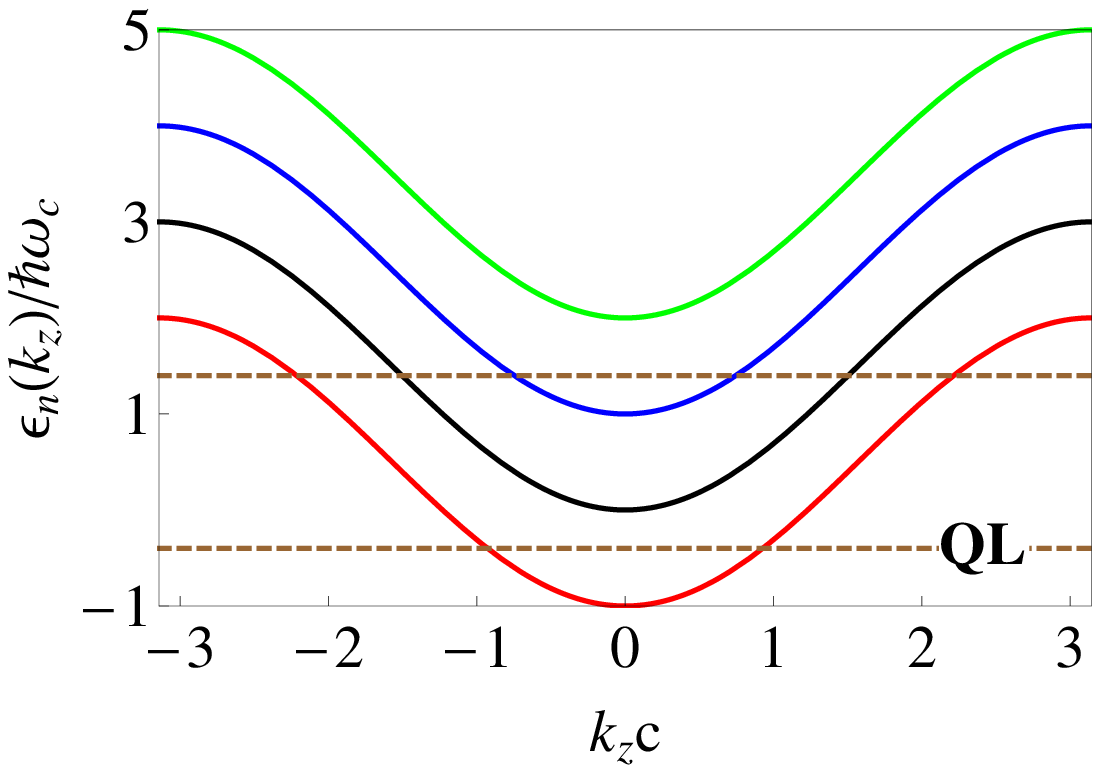}
\label{fig:subfig1a}
}
\subfigure[]{
\includegraphics[scale=0.65]{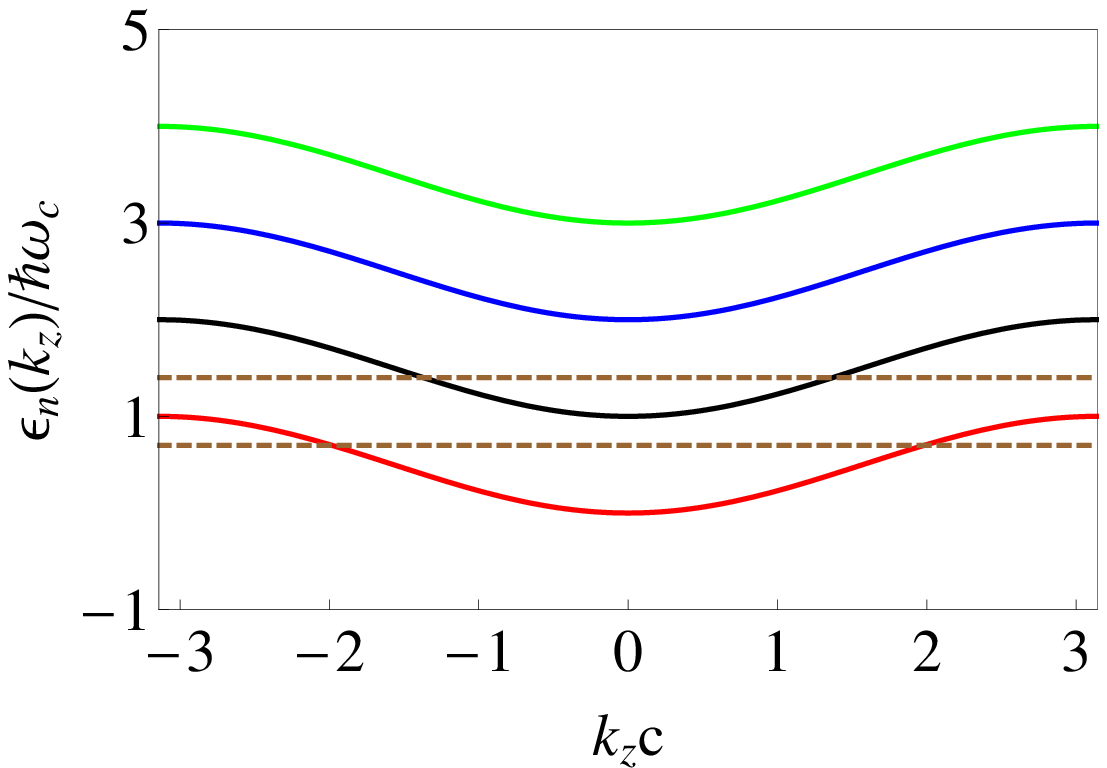}
\label{fig:subfig1b}
}
\label{fig:q2DLL}
\caption[]{(Color online)The dispersive LLs for a quasi-two dimensional metal [Eq.~(\ref{q2DLL})], when the magnetic field is applied along the $c$ axis. (a) $4t_z > \hbar \omega_c$ and (b) $4t_z< \hbar \omega_c$. For a high carrier density corresponding to Eq.~(\ref{pyrite}), the underlying Fermi surface is a corrugated cylinder and $\hbar \omega_c$ exceeds $4t_z$ before reaching the quantum limit. In this case, each LL is well separated as shown in panel (b) which gives rise to two Fermi points, and the system behaves effectively like the one in the quantum limit.}
\end{figure*}
\end{widetext}

\section{Quasi-two dimensional metals} \label{sec:QL}
In this subsection we consider the axial anomaly induced LMC of quasi-two dimensional metals. By quasi-two dimensional metals we are referring to layered materials where the inter-plane hopping matrix element is much smaller than its in-plane counterparts. Many interesting materials such PdCoO$_2$, PtCoO$_2$, graphite and $\alpha$-(BEDT)$_2$I$_3$ belong to this category. We will focus our discussion in Subsec.~\ref{delafossite} on PdCoO$_2$, which has a single sheet of Fermi surface~\cite{band3,arpes,Hicks,balicas}. In Subsec.~\ref{organic} we consider the qualitative agreement between our theory and the longitudinal magnetotransport measurements on $\alpha$-(BEDT)$_2$I$_3$. Towards the end we also provide a brief discussion of experiments on graphite.

\subsection{PdCoO$_2$ and PtCoO$_2$}\label{delafossite}For simplicity we consider the following Hamiltonian
\begin{equation}
H_{q2D}=\frac{\mathbf{p}^2_\perp}{2m_\perp}-2t_z \cos(k_z c),
\end{equation}
where $p_{\perp} = (p_x, p_y,0)$, $m_\perp$ is the effective mass in the $ab$ plane, and $c$ is the lattice spacing in the interlayer direction. When interlayer coupling is sufficiently weak in comparison to the energy scale of the in plane motion i. e.,
\begin{equation}
t_z < \frac{\pi \hbar^2 n c}{2m_\perp},\label{pyrite}
\end{equation}the underlying Fermi surface is a corrugated cylinder. This is the situation for PdCoO$_2$, PtCoO$_2$. In contrast for a stronger interlayer hopping such that
\begin{equation}t_z > \frac{\pi \hbar^2  n c}{2m_\perp},\label{graphite}\end{equation}the system possesses an ellipsoidal Fermi surface. In this case, one can expand the $\cos(k_zc)$ and obtain an effective parabolic dispersion like the one for an electron gas. Even though graphite has a more complicated band structure due to the presence of multiple atoms in a unit cell, which gives rise to a compensated semimetal phase with both electron and hole pockets~\cite{weiss,mclure}, some of the qualitative features can still be understood in terms of an electron gas with an ellipsoidal Fermi surface. In particular there are some interesting differences between PdCoO$_2$ and graphite in a tilted magnetic field, which we can easily understand on the basis of a simple Hamiltonian $H_{q2D}$.

When the magnetic field is applied along the $c$ axis, the dispersion relations for the LLs are described by
\begin{equation}
\epsilon_N(k_y,k_z)=\left(N+\frac{1}{2}\right)\hbar \omega_c -2t_z \cos(k_z c)\label{q2DLL},
\end{equation} where $\omega_c=eB/m_\perp$. The dispersive LLs for $4t_z> \hbar \omega_c$, are shown in Fig.~\ref{fig:subfig1a}, when the Fermi level can cross multiple LLs similar to the situation in an electron gas. In contrast, when $4t_z< \hbar \omega_c$, all the LLs are well separated, and the Fermi energy can intersect only one LL at a time, which is illustrated in Fig.~\ref{fig:subfig1b}. For a material like PdCoO$_2$, which satisfies Eq.~(\ref{pyrite}), $\hbar \omega_c$ has to exceed the scale $4t_z$ before the magnetic field can confine all the carriers into the LLL (true quantum limit). This is a unique situation when one almost mimics the quantum limit for a smaller field strength, where the axial anomaly due to a single LL contributes to the LMC. In this case, the inter LL scattering effects do not appear in the Boltzmann calculation (similar to the quantum limit), and
we only need to use the effective potential
\begin{eqnarray}
|U_{eff,N}(q_z)|^2&=&\int \frac{d^2q_\perp}{(2\pi)^2l^2_B}|U(q_\perp,q_z)|^2  \exp \left (-\frac{q^2_\perp l^2_B}{2}\right ) \nonumber \\
&&\times \left[L_N\left(\frac{q^2_\perp l^2_B}{2}\right)\right]^2,\label{UNqz}
\end{eqnarray}for the corresponding partially occupied LL. In this case for short range scattering it is possible to find a closed form expression for an arbitrary LL. For ionic scattering we can also obtain the corresponding expressions in closed form, but they become progressively lengthier with the LL index N. Some of these details can be found in the Appendix. When
$$2 \pi n cl^2_B \leq 1,$$ only the LLL is partially occupied, and the LMC in the quantum limit due to the short range scattering is given by
\begin{eqnarray}
\sigma^s(B)&=&\frac{4e^2t^2_z c^2}{hn^s_iU^2_0}\sin^2(2\pi^2ncl^2_B)\left(1+\frac{2a^2}{l^2_B}\right)\nonumber \\ && \times \exp[16\pi^2n^2l^4_Ba^2],\label{eqsigmaslayered}
\end{eqnarray} where we have used $k_{F,0}=2\pi^2nl^2_B$, and $v_{F,0}=2t_zc\sin(k_{F,0}c)/\hbar$. Due to the periodic dependence of the Fermi velocity on $B$, we have a nonmonotonic behavior of this contribution to the LMC, which initially increases with the magnetic field strength until $4\pi ncl^2_B<1$. For a stronger magnetic field, the LMC due to short range scattering decreases. When, $4\pi ncl^2_B \ll 1$, the $\sin^2$ term can be approximated as $\sim l^4_B$, and produces the same asymptotic behavior as in the case of an electron gas. Therefore, in the presence of only short range scattering, one may initially observe a negative LMR, followed by an upturn to a positive LMR, in the quantum limit. In contrast, the ionic scattering always leads to a strong positive LMC varying as $B^2$. Therefore, in the presence of both types of impurity scattering, the axial anomaly induced LMR follows the generic behavior illustrated in Fig.~\ref{LMR}.

\begin{widetext}
\begin{figure*}[tb]
\centering
\subfigure[]{
\includegraphics[scale=0.6]{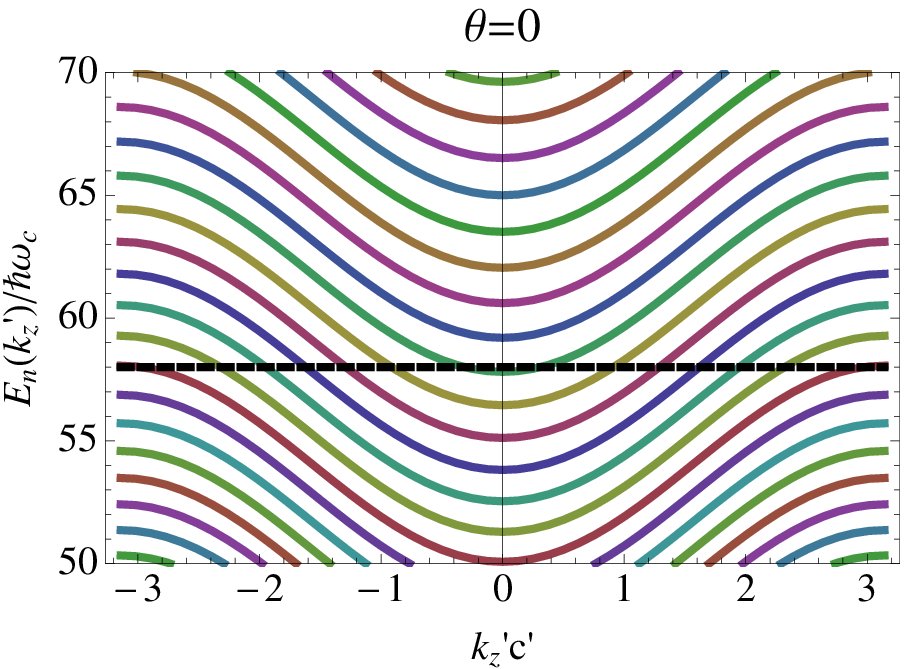}
\label{fig:subfig5a}
}
\subfigure[]{
\includegraphics[scale=0.6]{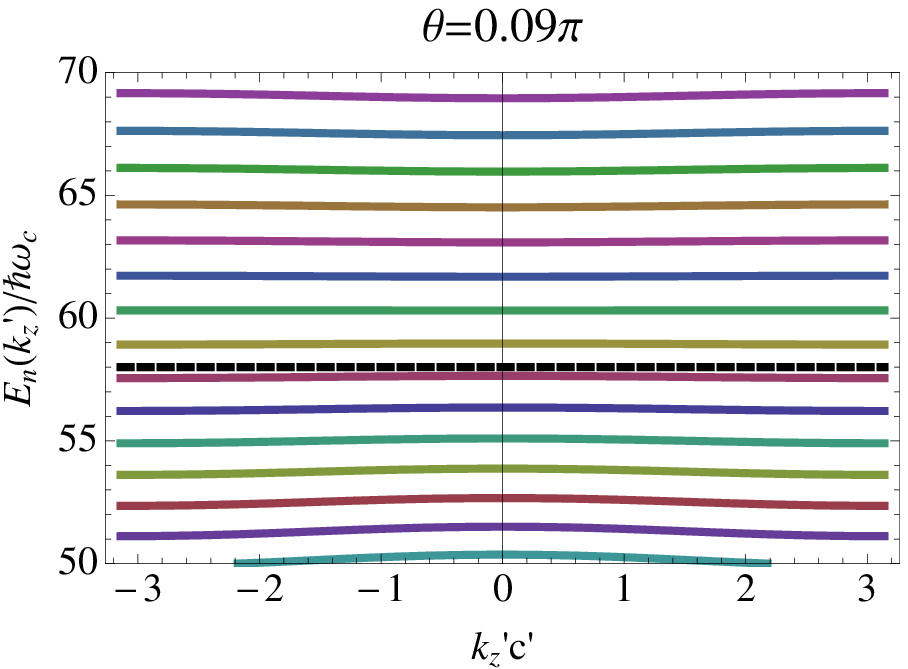}
\label{fig:subfig5b}
}
\subfigure[]{
\includegraphics[scale=0.6]{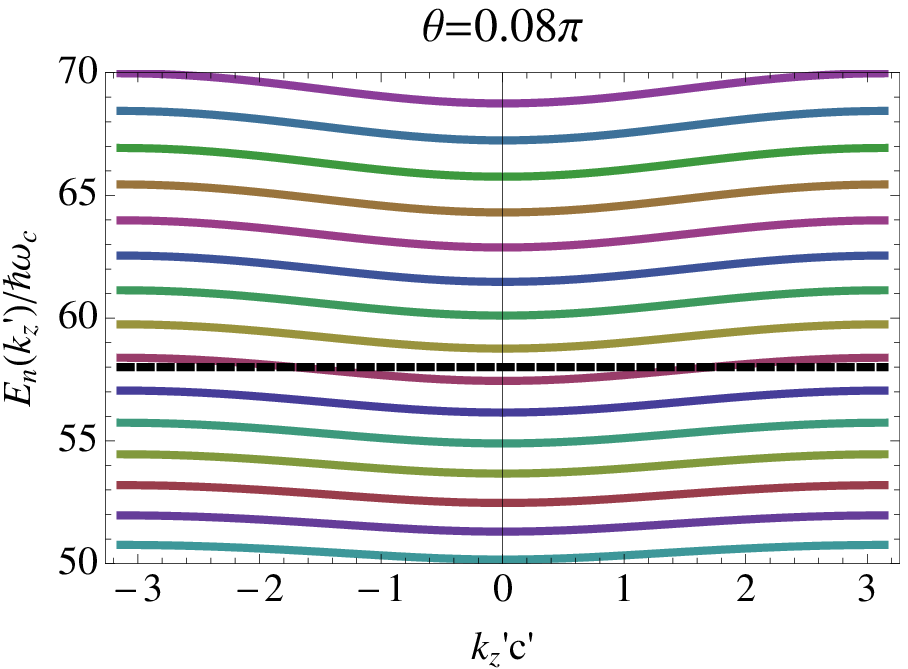}
\label{fig:subfig5c}
}
\label{fig:tilted}
\caption[]{(Color online)The dispersive LLs of a quasi-two dimensional metal with a high carrier density in a magnetic field (solid) applied at an angle $\theta$ with respect to the $c$ axis, as a function of $k_z'c'$ (defined in the text), for $t_z/(\hbar \omega_c)=2$. We are considering a large Fermi energy such that the Fermi level crosses only the LLs with large index $N$. For (a) $\theta=0$ there are a large number of Fermi points from multiple LLs. (b) For the same field strength and tilt angle $\theta =  0.09\pi$ the LLs with high index $N$ become completely flat and do not produce any Fermi point. Therefore the effective hopping strength along the field direction, for a large LL index vanishes, showing the existence of the Yamaji angle, independent of the magnitude of $t_z/(\hbar \omega_c)$. (c) For a small deviation away from the Yamaji angle, at $\theta=0.08\pi$ there is only one partially filled LL producing two Fermi points, similar to the situation in the quantum limit.}
\end{figure*}
\end{widetext}

In a material like PdCoO$_2$, to attain $4t_z<\hbar \omega_c$, one requires $B> 96$T.~\cite{balicas} Therefore, multiple partially occupied LLs contribute to the LMC for any experimentally accessible field strength. The calculation of LMC using Eq.~(\ref{sigmabN})is quite involved and it is not easy to unambiguously relate the observed negative LMR with the axial anomaly at a quantitative level. How can we then probe the large negative LMR due to the axial anomaly of a single LL? This can be achieved by placing the system in a tilted magnetic field. Due to the high carrier density in each layer, for any experimentally accessible steady field strength the Fermi level crosses only the LLs with large index $N$. When the field is tilted away from the $c$ axis, the effective interlayer hopping is diminished, and for certain tilt angles $\theta_Y$, also known as the Yamaji angles, the LLs with large $N$ become completely flat or nondispersive~\cite{Yamaji,Kurihara}, which can not produce any Fermi point. A slight deviation from $\theta_Y$ can restore a weak dispersion of these LLs along the applied field direction. Consequently, even for a moderate field strength we can have a single partially occupied LL, similar to the situation in Fig.~\ref{fig:subfig1b}. The original calculation of Yamaji~\cite{Yamaji} is a semiclassical one, and does not explicitly consider the Fermi points produced by the dispersive LL's, which play the most important role in our theory. Subsequently, the dispersion relations of the LLs have been calculated perturbatively in the limit~\cite{Kurihara} $t_z/(\hbar \omega_c \cos \theta)<<1$, where $\theta$ is the angle between the $c$ axis and the magnetic field $\mathbf{B}$. However, the experiments are performed in the opposite limit~\cite{balicas} $t_z/\hbar \omega_c > 1$. In addition, when $\theta$ becomes very small, the control parameter always becomes much bigger than unity. For this reason, we have numerically solved the relevant Schr{\"o}dinger equation with an arbitrary value of $t_z/\hbar \omega_c$, from which we have established the existence of $\theta_Y$ as a nonperturbative quantum mechanical effect.

We consider a tilted magnetic field ${\bf B} = (0,B\sin\theta,B\cos\theta)$ making an angle $\theta$ with the $c$ axis and choose the vector potential ${\bf A} = (0, xB\cos\theta,-xB\sin\theta)$. After the Peierl's substitution $\mathbf{k} \to \mathbf{k} -e\mathbf{A}/\hbar$, we arrive at the  following Hamiltonian
\begin{eqnarray}
H_{q2D} &=& -\frac{\hbar^2}{2m_\perp}\partial_x^2 + \frac{\hbar^2}{2m_\perp}(k_y - m\omega_c\cos\theta x)^2
\nonumber
\\
 &-&2t_z\cos(c[k_z+m\omega_c\sin\theta x]).
\end{eqnarray}
The momentum component along the field direction $k_z' = \sin\theta k_y + \cos\theta k_z$ and $k_y$ by itself are good quantum numbers. This becomes transparent, when we transform the $x$ coordinate to $x'=x-k_y/(m_{\perp}\omega_c\cos\theta)$, which leads to
\begin{eqnarray}
H_{q2D} &=& -\frac{\hbar^2}{2m_\perp}\partial_{x'}^2 + \frac{1}{2}m_\perp(\hbar\omega_c\cos\theta x')^2
\nonumber
\\
&-&2t_z\cos(c'[k_z'+m_{\perp}\omega_c\cos\theta\sin\theta x'])\label{tiltedLL},
\end{eqnarray}
where $c'=c/\cos\theta$. We numerically obtain the eigenvalues and the eigenstates of this Hamiltonian in Eq.~(\ref{tiltedLL}) after discretizing   $x^\prime$.

We consider a model parameter $t_z/(\hbar\omega_c)=2$, which lies well outside the perturbative regime. The Fig.~\ref{fig:subfig5a} corresponds to the field along $c$ axis, where many LLs cross the Fermi energy. The Fig.~\ref{fig:subfig5b} corresponds to the field tilted at the first Yamaji angle $\theta = 0.09 \pi$. At this angle, we clearly see the existence of non-dispersive LLs with high index $N$. For a small LL index $N$, the levels are still dispersive. This is the reason, why in contrast to the dense delafossite materials, a dilute system like graphite can not show the existence of Yamaji angle in AMRO measurements. It is important to note that there exist multiple Yamaji angles and within a perturbative treatment these are given by the zeros of a Bessel function~\cite{Yamaji,Kurihara}. For a small deviation from the first Yamaji angle, we find a single, weakly dispersive LL intersecting the Fermi energy, which is shown for $\theta=0.08 \pi$ in Fig.~\ref{fig:subfig5c}. This yields only two Fermi points similar to the situation in the quantum limit.

When the field is tilted, the measured magnetoresistance along the $c$ axis is a combination of the LMR along the field direction and the TMR in the plane perpendicular to $\mathbf{B}$. At the Yamaji angle, there are no dispersive LL with Fermi points, and we do not have any axial anomaly. For this reason, at the Yamaji angles, the magnetoresistance along the $c$ axis is completely determined by the positive TMR in the plane perpendicular to the applied $\mathbf{B}$. In contrast, for a small deviation from the Yamaji angle, the axial anomaly of a single dispersive LL will give rise to a LMR along the $\mathbf{B}$. In experiments on PdCoO$_2$, a small deviation from the $\theta_Y$ reveals the emergence of a negative component of the magnetoresistance measured along the interlayer direction. Therefore, axial anomaly in this material indeed leads to a negative LMR. More striking aspect is that for a sufficiently large $B$, the negative LMR actually subdues the very large positive TMR. At present we do not have any quantitative understanding of this effect. But, the observation certainly provides evidence for a very large magnitude of the anomaly induced negative LMR from a single partially filled LL, as anticipated in the quantum limit.

\subsection{$\alpha$-(BEDT-TTF)$_2$I$_3$}\label{organic}
$\alpha$-(BEDT-TTF)$_2$I$_3$ is a layered material, which under high pressure realizes a semimetallic state~\cite{Tajima1}. The intralayer dispersion is captured by two tilted Dirac cones and the system is half-filled~\cite{Kobayashi1,Kobayashi2}. Without the tilt, the in-plane dispersion is analogous to the quasiparticle spectrum of graphene. Here the band degeneracy is accidental, and do not occur at any high symmetry points. A negative LMR and a positive TMR have been concomitantly observed in this material for $B>0. 7T$~(Refs.~\onlinecite{Tajima2,Osada}). For understanding the salient features of the longitudinal magnetotransport measurements, we can ignore the tilt of the Dirac cone in the $ab$ plane as in Ref.~\onlinecite{Osada}. But, instead of a perturbative expansion of conductivity in interlayer hopping strength $t_z$, we retain the full dispersion along the $c$-axis.

We consider the following effective Hamiltonian
\begin{equation}
H_{or}=\hbar v \mathbf{k}_\perp \cdot \boldsymbol \tau \otimes \eta_3-2t_z \cos (k_z c),
\end{equation} where the Pauli matrices $\tau_j$ and $\eta_j$ respectively act on the sublattice and the valley indices. The dispersion relations for the conduction and the valence bands are given by
\begin{equation}
\epsilon_{\mathbf{k}}=\pm \hbar v k_\perp -2t_z \cos (k_z c).
\end{equation}  From the dispersion relation we can show that the density of states behaves as
\begin{equation}
D(\epsilon) \propto \frac{|\epsilon|}{c (\hbar v)^2}.
\end{equation} Therefore, for a system at half-filling the thermally excited carrier density (both electron and holes) will be $n_T \sim T^2$, in agreement with experiments. Due to the very low carrier density the conduction and the valence bands give rise to electron and hole pockets respectively centered around $k_z=0$ and $k_z=\pi$. At half-filling $(\epsilon_F=0)$, the volume of the electron and hole pockets (elongated along the $c$ axis) are equal, and they both end at $k_z=\pm \pi/2$. Thus, the system behaves as a compensated semimetal.

In the presence of a quantizing magnetic field, the LL dispersion around each valley is given by
\begin{equation}
\epsilon_{N\neq 0, k_z, s}= \pm \sqrt{2N} \frac{\hbar v}{l_B}-2t_z \cos(k_z c) - s \; \Delta_Z,
\end{equation} where $s=\pm 1$ denote the spin projections and $\Delta_Z= g\mu_B B/2$. In contrast the dispersion for the zeroth LL is captured by
\begin{equation}
\epsilon_{N=0, k_z,s}=-2t_z \cos(k_z c) - s \; \Delta_Z.
\end{equation} When
\begin{equation}
\sqrt{2}\frac{\hbar v}{l_B}> (4t+2\Delta_Z),
\end{equation} only the LLL can be partially filled. At half-filling, the spin down and the spin up LLL become completely empty and filled respectively if the Zeeman splitting exceeds the band width $2t_z$, and the LMC vanishes (system behaves as a band insulator). Therefore a finite LMC in the quantum limit can only occur for $2t_z> \Delta_Z$. The Fermi wavenumbers in the LLL are given by
\begin{equation}
k_{F,s}=\frac{\pi}{2c}+ s \; \frac{1}{c} \arcsin \left(\frac{\Delta_Z}{2t_z}\right).
\end{equation} The LMC due to the short range scattering is now given by
\begin{eqnarray}
\sigma^s(B)&=&\frac{8e^2t^2_z c^2}{hn^s_iU^2_0}\left[1+\frac{2a^2}{l^2_B}\right]\left[1-\frac{\Delta^2_Z}{4t^2_z}\right]\times  \nonumber \\ &&(e^{4k^2_{F,+}a^2}+e^{4k^2_{F,-}a^2}) \Theta(2t_z-\Delta_Z),
\end{eqnarray} where $\Theta(x)$ is the Heaviside step function. Therefore, zero range point impurities can only cause a positive LMR. In contrast, for weak enough magnetic fields the Gaussian impurities can lead to a $B$ linear positive LMC, and in this regime the negative LMR will vary as $\rho \sim 1/(B^\ast + B)$, which has been observed in Ref.~\onlinecite{Tajima2}. In the presence of ionic impurities, the LMR can initially decrease as $1/B^2$ until the suppression due to the Zeeman splitting becomes strong enough to cause a positive LMR. Therefore, the overall profile for the LMR is similar to the one in Fig.~\ref{LMR}, with the exception that the LMR finally goes to infinity for $\Delta_Z>2t_z$.

Before leaving the topic of quasi two dimensional metals, we will also briefly mention the experimental situation in graphite. Due to the very low carrier density, it is easy to attain the quantum limit with a moderately strong magnetic field $B \sim 7.5T$ along the $c$ axis~\cite{behnia1}. At a stronger magnetic field graphite suffers an instability towards a charge density wave formation with a wavevector $2k_{F,0}(B,n)$ around $B \sim 25T$~(Refs.~\onlinecite{fukuyama, behnia1}). For $9 T< B<B_c(T)$, where $B_c(T)$ is the threshold for the first instability of the ground state, the magnetotransport data shows concomitant negative LMR and positive TMR (see Fig. 1 of Ref.~\onlinecite{behnia1}). This range becomes larger with increasing temperature. A more sophisticated treatment that accounts for electronic interaction is required for addressing the magnetotransport when $B>B_c(T)$.

\section{Weyl semimetal}\label{subsec:Weyl}
WSM is a gapless state where the nondegenerate conduction and the valence bands touch linearly at isolated points in the Brillouin zone. For this reason, the touching points act as the monopole and the antimonopole of the U(1) Berry flux, and lead to many intriguing transport and optical properties~\cite{Zyuzin3,Grushin,GoswamiTewari}. One must break time reversal or inversion symmetry for realizing a WSM. There are many theoretical proposals for realizing a WSM~\cite{vishwanath,xu,Burkov1,Burkov2,Zyuzin1,Cho,Meng,Gong,Sau,Das,ganeshan}. But, a concrete experimental example of a noncentrosymmetric Weyl semimetal phase has only been found very recently in TaAs~\cite{TaAs1,TaAs2,TaAs3,TaAs4}, and the band structure calculations also predict the existence of such a phase in TaP, NbAs and NbP~\cite{TaAs1,TaAs2}. There is also some preliminary evidence for a WSM in $\beta$-Ag$_2$Se~\cite{AgSe}. Given that a massless Dirac fermion is made up of two Weyl fermions of opposite chirality, which are located at the same point in the Brillouin zone, and a Zeeman coupling explicitly lifting the spin degeneracy can convert a DSM into a WSM, it is important to first understand the magnetotransport for the WSM.

For this reason, we consider the following toy model on a tetragonal lattice
\begin{eqnarray}
&&H_{WSM}=2t_{\perp,1} [\sin (k_xa) \sigma_1+\sin (k_ya) \sigma_2]-[2t_{\perp,2} \{
 2- \nonumber \\ && \cos (k_xa)-\cos (k_ya)\}  +2t_{z}\cos (k_zc)-\Delta]\sigma_3,
\end{eqnarray} to describe a WSM, where $a$ and $c$ are respectively the lattice spacings in the $ab$ plane and along the $c$ axis. For $|\Delta/t_{z}|<1$, conduction and valence bands touch at
\begin{equation}
\mathbf{k}=\left(0,0,\pm \frac{1}{c} \arccos(\Delta/2t_{z})\right),
\end{equation}
which are the respective locations of the left and the right handed Weyl fermions. The tight binding parameter $t_{\perp,2}$ prohibits the existence of additional gapless points for $k_x \neq 0$ and $k_y \neq 0$.
\begin{figure}[htb]
\includegraphics[width=8cm,height=6cm]{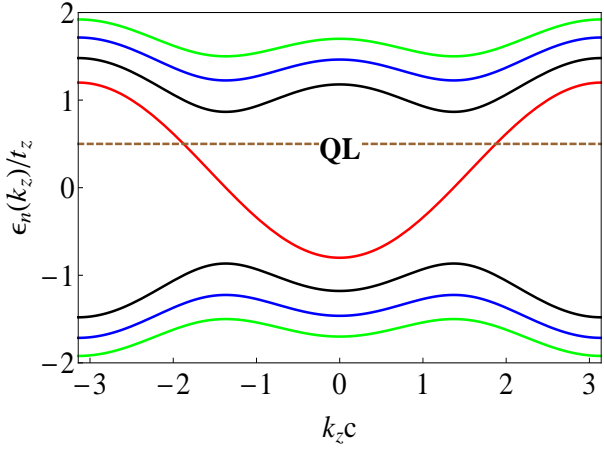}
\caption{(Color online) The dispersing LLs of a three dimensional WSM. We have chosen the following parameters $\Delta/t_{z}=0.2$, $t_{\perp,1}/t_{z}=20$, and $B=2$T. The Landau levels with N=0, 1, 2 and 3 are respectively represented with the colors red, black, blue and green.}
\label{WeylLandaulevels}
\end{figure}
For studying the LLs of the Weyl fermions, we linearize $H_2$ in $k_x$ and $k_y$, and perform the Peierls substitution $\mathbf{k} \to \mathbf{k} -e\mathbf{A}/\hbar$. However, we do not linearize in $k_z$, for illustrating how a dispersing LL behaves in the entire first Brillouin zone $-\pi <k_zc<\pi$ along the $c$-axis. In the Landau gauge $\mathbf{A}=(0,Bx,0)$, the Hamiltonian becomes
\begin{eqnarray}
H_{WSM} &\approx& \hbar v_{\perp}[-i\partial_x \sigma_1+(k_y-eBx/\hbar) \sigma_2]
\nonumber
\\
 &+&-[2t_{z}\cos (k_zc)  -\Delta ]\sigma_3,
\end{eqnarray}
where $v_\perp=2t_{\perp,1}a/\hbar$. The dispersion relations for the Landau levels are given by
\begin{eqnarray}
\epsilon_{N=0}(k_y,k_z)&=&-2t_z \cos(k_zc) + \Delta, \\
\epsilon_{N\neq0}(k_y,k_z,\alpha)&=& \mathrm{sgn}(\alpha) \bigg[(2t_z \cos(k_zc) -\Delta)^2 \nonumber \\ &&+ 2N \frac{\hbar^2 v^2_\perp}{l^2_B}\bigg]^{1/2}\label{WeylhigherLandau},
\end{eqnarray} where $\alpha=\pm 1$. The dispersive LLs for N=0, 1, 2 and 3 are showed in Fig.~\ref{WeylLandaulevels} for a particular choice of tight binding parameters. The LLL crosses the zero energy at $k_zc=\pm \arccos(\Delta/2t_z)$, and the two minima of the higher Landau levels also occur at the same locations in the Brillouin zone. An explicit consideration of Zeeman coupling, only shifts these minima and the Fermi points of the LLL to new locations, as long as $(\Delta +\Delta_Z(B))<2t_z$, where $\Delta_Z(B)$ is the Zeeman splitting. However, this does not introduce any qualitative change to our conclusions regarding the magnetoconductivity in the quantum limit. Both at and away from the half-filling the system behaves as a metal and we can always find a set of Fermi points. If $(\Delta +\Delta_Z(B))>2t_z$, the minima of the $N \neq 0 $ , $\alpha=1$ and $N=0$ LLs occur at $k_z=0$ At the same time the maxima of $N \neq 0 $ , $\alpha=-1$ LLs are also located at $k_z=0$. At half-filling the $N=0$ LL is still partially filled with $k_{F,0}=\pi/2$, and the system acts as a metal. We also note that, for the $N \neq 0$ LLs, both upper and lower components of the spinor wave function are nonzero and they correspond to $\varphi_{N}$ and $\varphi_{N-1}$, where $\varphi_{N}$ is the wave function for $N$-th Landau level of an electron gas. In contrast, for the LLL the lower component of the spinor wavefunction vanishes and the upper component is $\varphi_0$.

When $\sqrt{2}\hbar v_\perp/l_B>(2t_z+\Delta)$, both $N=1$ LLs corresponding to $\alpha=\pm 1$ remain well separated from the LLL. This is the simplest situation for analyzing the LMC. For this relativistic model the Fermi wavenumber is given by
\begin{equation}
k_{F,0}(B)=\frac{\pi}{2c}+ 2\pi^2 l^2_B \delta ,
\end{equation}where $\delta$ measures the carrier density as the deviation from the half-filling. In particular when,
\begin{equation}
0<l^2_B \leq \frac{1}{4\pi c \delta } < \frac{2 (\hbar v_F)^2}{(\Delta+2t_z)^2},
\end{equation} only the lowest Landau level remains partially occupied. In this situation, the calculation of LMC is identical to the one for the layered metal presented in the previous section. The LMC due to the ionic scattering still varies as $\sim B^2$. In contrast, the LMC due to the short range scattering is given by \begin{eqnarray}
\sigma^s(B)&=&\frac{4e^2t^2_z c^2}{hn^s_iU^2_0}\cos^2(2\pi^2 l^2_B c \delta)\left(1+\frac{2a^2}{l^2_B}\right)\nonumber \\ && \times \exp \left[\frac{\pi^2a^2}{c^2}\left(1+ 4\pi l^2_B c \delta \right)^2\right].
\end{eqnarray} At half-filling, the zero range point impurities lead to a constant LMC, whereas the neutral Gaussian impurities give rise to a $B$-linear LMC (when $l^2_B<2a^2$). For a system away from the half-filling, an increasing magnetic field strength, asymptotically brings $k_{F,0}$ closer to the half-filled value $\pi/2c$. Therefore, for a finite $\delta$, the LMC due to the zero range point impurities grows as $\cos^2(B_0/(2B))$ towards saturation. In contrast, the finite range Gaussian impurities still lead to a $B$ linear LMC (as the argument of the exponential saturates to $\pi^2 a^2/c^2$). In the presence of both ionic and short range impurity scattering, the LMR can initially decrease as $B^{-2}$ before crossing over to the short range impurity dominated behavior. Therefore, for a pure WSM (with no carriers coming from additional parabolic bands) the LMR in the quantum limit can show upturn to become positive only due to an insulating ground state caused by the electronic interaction. But, the existence of additional parabolic bands will make the LMR to follow the generic behavior illustrated in Fig.~\ref{LMR} even without any broken symmetry phase. For $(2t_z-\Delta)<\sqrt{2}\hbar v_\perp/l_B$, the LLs with $\alpha=-1$ remain completely filled, and do not contribute to the conductivity. The condition for quantum limit is then determined by
\begin{equation}
2t_z \sin(2\pi^2 l^2_B c \delta)+\Delta< \frac{\sqrt{2} \hbar v_\perp}{l_B},
\end{equation} and our conclusions remain unaltered. When $(2t_z-\Delta)>\sqrt{2}\hbar v_\perp/l_B$, LLs with $\alpha=-1$ start to become partially filled. Such a situation can be addressed by using Eq.~(\ref{sigmabN}) and a numerical evaluation of the transport lifetimes $\tau_{tr,N}$.

Before moving on to the discussion for other physical systems, we point out the essential difference between our results and the ones found in the existing literature. To the best of our knowledge, the magnetic field dependence of the velocity $v_{F,0}$ and the transport lifetime $\tau_{b,0}(B)$ have been ignored in previous works~\cite{aji,son,kim1,ong,shovkovy}. In these past studies
the velocity $v_{F,0}$ has been taken as the Fermi velocity of the Weyl excitations in the absence of a magnetic field, and $\tau_{b,0}$ has been treated as a phenomenological constant (by ignoring the matrix elements of the LLL wavefunction). From this one obtains a $B$ linear magnetoconductivity (following Eq.~\ref{sigmab0}). Through our analysis we have showed that both $v_{F,0}(B)$ and $\tau_{b,0}(B)$ have nontrivial magnetic field dependence, which in the quantum limit can not be ignored on any physical ground.

For weak magnetic field strengths, the axial anomaly induced negative LMR of a WSM can also be understood from the semiclassical calculations of Son and Spivak.~\cite{son} This approach is valid in the absence of Landau quantization, i.e., $\omega_c \tau <<1$. In this case, the axial anomaly of Weyl fermion appears in the semiclassical equations of motion (in the plane wave or Bloch band basis) through Berry curvature. In contrast, our calculations are performed in the strong magnetic field regime $\omega_c \tau >>1$ (particularly in the quantum limit) in the spirit of original proposal of Nielsen and Ninomiya. These two theories can as such be considered complimentary since they are developed for different magnetic field regimes.

However, there is a common feature in both theories. Son and Spivak do not explicitly compute the transport lifetime (in contrast to our work where we explicitly calculate the transport scattering times in the system). Rather they made an important and tacit assumption, that the intervalley  scattering time (between Weyl points of opposite chirality) is considerably larger than the intravalley scattering time and the latter determines the transport lifetime. In the absence of a magnetic field intravalley scattering completely dominate the resistivity (i.e. intervalley scattering is unimportant in the system and can be ignored). This can only happen for long range impurity potential, since short range point impurities give equal intra and intervalley scattering rates. This requirement on the nature of the relaxation mechanism in the work of Son and Spivak is precisely what we have demonstrated from a Landau level based calculation, without making any tacit assumption regarding the transport lifetime. We explicitly show in our work that long-range impurity potential generically gives rise to negative magnetoresistance (which is what Son and Spivak find in their semiclassical treatment) whereas short-range impurity scattering (not considered by Son and Spivak) by contrast gives rise to positive magnetoresistance.  Thus, it seems that the finding of Son and Spivak is a special restricted case of our general theory where we find that the magnetoconductivity is positive (negative) for long-range (short-range) disorder potential.

An important question is whether one can smoothly interpolate between these two (i.e. Landau-quantized strong-field and semiclassical weak-field) limits. The semiclassical methods cannot capture the physics of the strong field limit (as soon as quantum oscillations come into play), as it is oblivious to the Landau level formation. In general it is quite challenging, but perhaps not impossible, to obtain such an interpolation starting from the Landau level basis. However, the interpolation between the weak-field semiclassical regime and the strong-field Landau quantized regime is outside the scope of the Boltzmann transport theory used in both our and Son-Spivak work.

If we consider TaAs as a typical example of WSM, magnetotransport experiments~\cite{TaAs4} suggest that the quantum limit is achieved around $B \sim 8T$, which is reasonable for a carrier density $n \sim 2.65 \times 10^{17}$ cm$^{-3}$. Therefore, the magnetic field range $5 T< B<9T$ for which a negative LMR is observed in this material in Refs.~\onlinecite{XHuang,CZhang} does not belong to the semiclassical regime considered by Son and Spivak. Nevertheless, the LMC in this regime shows $B^2$ dependence, and a subsequent upturn towards positive value. From the comparison of the quantum lifetime (extracted from Dingle temperature) and the measured transport lifetime, it is found that the transport lifetime is almost $100$ times larger than the quantum lifetime.~\cite{TaAs4} This essentially points towards the dominance of the forward scattering mechanism due to the long range impurities. Therefore, the observations in this material are consistent with our theoretical predictions.

We mention, however, a direct experimental way of verifying our prediction of an impurity-potential-range induced crossover between positive and negative LMR in WSM systems.  This can be easily done by systematically introducing short-range (or long-range) disorder in the materials either through radiation damage or by actually incorporating atomic level impurities (or charged impurities) in the system, and measuring transport properties of different samples using different levels of controlled impurity incorporation.  Such an experimental technique was instrumental in settling the importance of long-range versus short-range disorder in graphene transport properties~\cite{EHHwang}. We urge similar experiments in WSM systems to investigate the importance of long- versus short-range disorder in producing positive versus negative magnetoconductance respectively as predicted in our theory.

Recently a very large negative LMR has also been observed in the WSM phase of TaP between 5 T and 14 T (see Ref.~\onlinecite{Shekhar}). In this range of magnetic field strengths, the LMC behaves as $B^2$, which is precisely consistent with our theory (with predominant long-range ionic disorder scattering). In contrast to the other WSM TaAs where the LMR shows saturation around 9T and a subsequent upturn, no such saturation has been observed in TaP up to 14 T. Despite possessing similar band structures, these two materials do have considerably different saturation field strengths ($B^\ast$ for TaP is yet to be determined from measurements at $B>14$ T). This further vindicates our result that the saturation or crossover field strength $B^\ast$ is nonuniversal. As discussed above, our theory predicts that the introduction of controlled amount of short-range disorder in TaP should suppress the crossover field for the change from negative to positive LMR.

\section{Dirac semimetal: two cones}\label{subsec:2Dirac}
When two Kramers degenerate conduction and valence bands linearly touch at isolated points in the Brillouin zone we obtain a DSM. The quasiparticles are then described by the four component massless Dirac equations. The DSM phase with two Dirac cones~\cite{Dai1,Dai2} located at $(0, 0, \pm k_0)$, has been experimentally realized in Cd$_3$As$_2$~\cite{Zahid1,Cava1,ZX1}, Na$_3$Bi.~\cite{ZX2,Zahid2} The pertinent Hamiltonian is built out of two copies of Weyl Hamiltonian $H_2$ in the following way
\begin{eqnarray}
&&H_{D,2}=H_{WSM} \otimes \tau_3.
 \end{eqnarray} For this reason, the quasiparticles' velocities along the $c$ axis at two cones have opposite signs. If we linearize the dispersion around the Dirac points we can write the following low energy Hamiltonian
\begin{equation}
H_{D,2} \approx \hbar v_z k_z \alpha_z \otimes \tau_3 + \hbar v_\perp (k_x \alpha_x+k_y \alpha_y) \otimes \tau_3,
\end{equation} where $\alpha_j$'s are anticommutinng $4\times 4$ Dirac matrices. The two Dirac cones have opposite chiral charge, which corresponds to the product of the velocities along three directions.  For a moment let us assume that the Fermi energy is pinned at the Dirac points. Then a relativistic field theory based calculation of the axial anomaly shows that
\begin{eqnarray}
\partial_\mu j_{\mu,5,1}+\partial_\mu j_{\mu,5,2}=0, \\
\partial_\mu j_{\mu,5,1}-\partial_\mu j_{\mu,5,2}=\frac{e^2}{\pi^2 \hbar} \; \mathbf{E}\cdot\mathbf{B},
\end{eqnarray} where $1$ and $2$ denote the two flavors of the Dirac fermions. Therefore, the net axial anomaly (flavor singlet) cancels due to the presence of the doublet. In such a situation one may naively think that the axial anomaly does not contribute to the transport. However, a LL based calculation will clearly demonstrate that such a conclusion is invalid for determining the LMC.

\begin{figure}[htb]
\includegraphics[width=8cm,height=6cm]{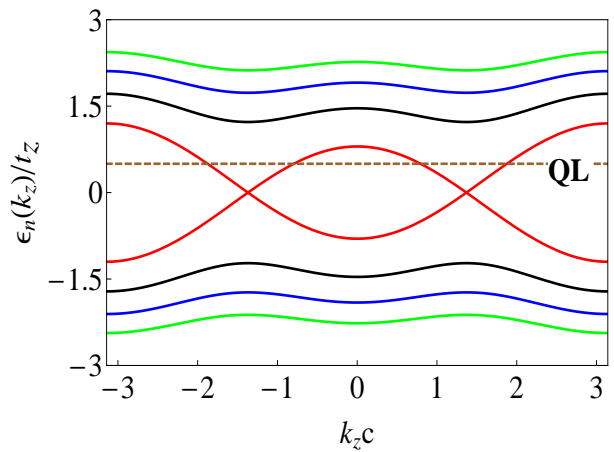}
\caption{(Color online)The dispersing LLs of a DSM with two Dirac cones. We have chosen the following parameters $\Delta/t_{z}=0.2$, $t_{\perp,1}/t_{z}=20$, and $B=2$T. The LLS with N=0,1,2 and 3 are respectively represented with the colors red, black, blue and green. In contrast to a WSM, the $N \neq 0$ LLs are two fold degenerate, and there are two N=0 levels. In the quantum limit as denoted by the dashed brown line, the carriers in the two levels are respectively the electrons and the holes.}
\label{DiracLandaulevels}
\end{figure}
The LLs with $N \neq 0$ are now two fold degenerate, with the dispersion relation still given by Eq.~(\ref{WeylhigherLandau}). A new feature is the presence of two $N=0$ LLs with
\begin{equation}
\epsilon_{N=0}(k_y,k_z,\alpha)=\mathrm{sgn}(\alpha)[-2t_z \cos(k_zc) + \Delta].
\end{equation} Notice that the two $N=0$ LLs levels have opposite curvatures. The dispersing LLs for a DSM with two cones are showed in Fig.~\ref{DiracLandaulevels}. This type of LL structure after accounting for a shift caused by the Zeeman splitting becomes consistent with the recent LL spectroscopy measurements performed on Cd$_3$As$_2$.~\cite{yazdani} The magnetoconductivity in the quantum limit is determined by the electron like carriers from the $N=0, \alpha=1$ level and the hole like carriers from the $N=0, \alpha=-1$ level. For this reason each level follows a formula similar to the ones for a WSM, where we need to use the appropriate electron and hole like carrier densities. We also emphasize that there is no impurity induced matrix element between these levels. For a net carrier density $\delta$ measured with respect to the half-filling, the Fermi wavenumbers of these LLs are
\begin{equation}
k_{F,\alpha}=\frac{1}{c}\arccos\left(\frac{\Delta}{2t_z \cos(\pi^2 l^2_B c\delta)}\right) + \alpha \; \pi^2 l^2_B \delta,
\end{equation} and an increasing magnetic field tends to bring the wavenumbers close to the corresponding values at half-filling. After substituting these $k_{F,\alpha}$ in Eq.~(\ref{eqsigmaslayered}) we can obtain the explicit expressions for the LMC due to each LL. We find that the LMC due to the axial anomaly is finite even if the Fermi energy is at the Dirac points (as assumed for a relativistic vacuum). The finite range Gaussian scattering gives rise to a $B$-linear positive LMC and zero range scattering only gives a constant LMC. If the long-range ionic scattering is the only source of relaxation, the LMC will again grow as $B^2$.

These observations remain true when Zeeman spin splitting can be ignored. For a sufficiently strong magnetic field, the Zeeman splitting modifies the dispersion
of the $N=0$ levels to
\begin{equation}
\epsilon_{N=0}(k_y,k_z,\alpha)=\mathrm{sgn}(\alpha)[-2t_z \cos(k_zc) + \Delta +\Delta_Z(B)].
\end{equation} If, $2t_z<(\Delta+\Delta_Z(B))$, there is a gap between the two LLs and the half-filled system behaves as an insulator. For finding $k_{F,\alpha}$, we need to make the modification $\Delta \to \Delta +\Delta_Z(B)$. Therefore, the Zeeman splitting has the tendency to reduce the size of the Fermi wavenumbers, and the LMC. The asymptotic form for the suppression of the LMC due to the Zeeman splitting behaves as
$$\left[1-\frac{(\Delta+\Delta_Z(B))^2}{4 t_z^2}\right]\Theta(2t_z-\Delta-\Delta_Z(B)),$$ where $\Theta$ is the Heaviside step function. When the transport is dominated by ionic scattering, we initially expect a positive LMC varying as $B^2$, followed by its eventual upturn due to the Zeeman splitting effects. Therefore, the qualitative profile of the LMR will be similar to the one in Fig.~\ref{LMR}, till $\Delta_Z$ becomes very large to cause an insulating behavior.

The quantum limit in Cd$_3$As$_2$ is achieved around $B_0 \sim 43 T$.~\cite{ong,zhao} This is in qualitative agreement with a rough estimate of the quantum limit using $l_{B}<n^{-1/3}$. If the density $n \approx 1.9 \times 10^{24} \; m^{-3}$ is taken from the ARPES measurements~\cite{Zahid1,Cava1,ZX1}, one obtains $B_0 \sim 10.3 T$. In contrast, if the density estimated from the the Hall coefficient is chosen, which can be a factor of $10$ times larger than the previous one~\cite{ong}, $B_0$ is found in the range of $18 T$ to $41 T$. At present there is no published LMR data in the quantum limit. However, a small negative LMR component has been observed on top of a very large positive background even for much smaller field strength ($B \leq 10 T$)~\cite{ong}. Beginning around $B \sim 2.5 T$, the average value of the LMR shows a downturn due to the negative component. This quickly goes away if the angle between the applied current and the magnetic field deviates more than $4^{\circ}$. As a material, Cd$_3$As$_2$ belongs to the family of gapless semiconductors, and transport at low temperatures will be naturally dominated by ionic impurity scattering. Therefore, we predict that a systematic study of the LMR in Cd$_3$As$_2$ and related materials up to the quantum limit will reveal a $B^2$ dependence of the positive LMC and its eventual downturn caused by the neutral impurities (see Fig.~\ref{LMR}) and the Zeeman spin splitting.

Recently, Ref.~\onlinecite{XiongOng} has reported axial anomaly induced negative LMR in Na$_3$Bi, for which quantum limit is achieved around $B \sim 4-6T$. The ratio between the intervalley scattering lifetime and the conventional transport lifetime for this material has been estimated to be around $40-60$, which suggests the underlying relaxation mechanism due to the long range impurities. It has also been reported that the LMC varies as $B^2$ in a wide range of magnetic field strengths $1T < B< 35 T$. An increasing positive LMR background has been ascribed to a small misalignment between the current and the magnetic field. The results in the weak field limit can be understood within a semiclassical theory~\cite{son}. But, in the broad range of magnetic field strengths $4 T < B< 35 T$ (quantum limit) the semiclassical methods become invalid. The experimental results clearly corroborate our main finding that in the quantum limit long range ionic scattering gives rise to LMC $\propto B^2$.

\section{Dirac semimetals: odd number of cones}\label{subsec:1Dirac}
A DSM with an odd number of band touching points or flavors does appear at the transition between the strong $Z_2$ topological and trivial insulators~\cite{Goswami1}. Various examples of such materials have already been presented in the Sec.~\ref{sec:Intro}. A representative massless Dirac Hamiltonian for a tetragonal crystal with single cone is described by
\begin{eqnarray}
&&H_{D,1}=2t_{\perp,1}[\sin(k_xa)\alpha_1+\sin(k_ya)\alpha_2]+2t_z \sin (k_zc) \alpha_3 \nonumber \\
&&+ 2t^\prime [3-\cos(k_xa)-\cos(k_ya)-\cos(k_zc) ]\beta,
\end{eqnarray} where $\alpha_j=\sigma_j \otimes \tau_3$ and $\beta=\sigma_0 \otimes \tau_1$ are four anticommuting Dirac matrices in the chiral representation. The Hamiltonian anticommutes with $\beta \gamma_5$, where $\gamma_5=\sigma_0 \otimes \tau_3$. The single Dirac cone is located at the $\Gamma$ point $\mathbf{k}=(0,0,0)$. The Wilson mass term $\propto t^\prime$ is essential to prohibit additional Dirac cones at other high symmetry points located at $(\pi, \pi, \pi)$, $(\pi, \pi,0)$, $(\pi, 0,\pi)$, $(0,\pi, \pi)$, $(\pi, 0,0)$, $(0,\pi, 0)$, $(0,0,\pi)$. In the continuum limit this term causes a $k$ dependent hybridization between the right and the left handed Weyl fermions according to
\begin{equation}
H_{D,1} \approx \hbar \sum_{j=1}^{3} v_j k_j \alpha_j+ \beta \left[\frac{\hbar^2k^2_\perp}{2m_\perp}+\frac{\hbar^2k^2_z}{2m_z}\right],
\end{equation}where $v_\perp=2ta/\hbar$, $v_z=2tc/\hbar$, $m_\perp=\hbar^2/(2t^\prime a^2)$, $m_z=\hbar^2/(2t^\prime c^2)$ . In the presence of the momentum dependent Dirac mass term, $[H_4,\gamma_5] \neq 0$, and the continuous chiral symmetry with respect to $\gamma_5$ is absent. This is reduced to a discrete $Z_2$ chiral symmetry captured through the spectral symmetry criterion $\{H_4, \beta \gamma_5 \}=0$.
\begin{figure}[htb]
\includegraphics[width=8cm,height=6cm]{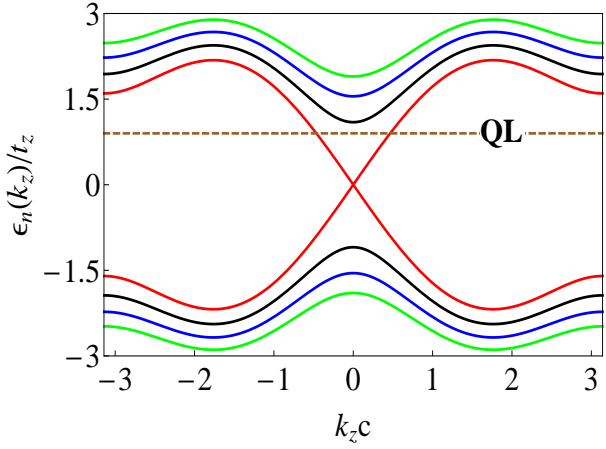}
\caption{(Color online)The dispersing LLs of a DSM with a single Dirac cone as determined from Eq.~(\ref{Dirac1LL}). We have chosen the following parameters $t_{\perp,1}/t_{z}=20$, $t^{\prime}/t=0.4$, and $B=2$T. The LLs with N=0, 1, 2 and 3 are respectively represented with the colors red, black, blue and green. The Landau levels with $N \neq 0$ are two fold degenerate. In contrast $N=0$ levels are nondegenerate.}
\label{Dirac1Landaulevels}
\end{figure} For the calculations of the Landau levels, if we ignore the higher gradient term such as $\beta \; (\mathbf{k}_\perp-e\mathbf{A}_\perp)^2/2m_\perp$,
the dispersion relations are given by
\begin{eqnarray}
\epsilon_{N}(k_y,k_z,\alpha)&=& \mathrm{sgn}(\alpha) \bigg[t^2\sin^2 (k_zc)+M^2(k_zc) \nonumber \\ &&+ 2n \frac{\hbar^2 v^2_\perp}{l^2_B}\bigg]^{1/2}\label{Dirac1LL},
\end{eqnarray} where $M(k_zc)=2t^\prime[1-\cos(k_zc)]$. In contrast to the $N=0$ level, all the higher Landau levels are two fold degenerate. The LLs described by Eq.~(\ref{Dirac1LL}) are illustrated in Fig.~\ref{Dirac1Landaulevels}.

An explicit consideration of $\beta \; (\mathbf{k}_\perp-e\mathbf{A}_\perp)^2/2m_\perp$ leads to a gap between the two $N=0$ levels, according to
\begin{eqnarray}
&&\epsilon_{N=0}(k_y,k_z,\alpha)= \nonumber \\ &&\mathrm{sgn}(\alpha) \bigg[t^2\sin^2 (k_zc)+\left[M(k_zc)+ \frac{\hbar \omega_{c,\perp }}{2}\right]^2\bigg]^{1/2},
\end{eqnarray} where $\omega_{c,\perp}=eB/m_\perp$. The size of the gap is $\hbar \omega_{c,\perp }$, and strictly speaking the half-filled DSM with odd number of cones is an insulator. The ratio of this gap and the bandwidth is roughly estimated to be $10^{-5} B \; t^{\prime}/t$, and can be neglected for all practical purpose. The formula for $N \neq 0$ are lengthy and not illuminating. For this reason they are not presented here.

So far, we have not considered an explicit Zeeman coupling. Usually, the semiconducting materials have very large and anisotropic $g$ tensors. In addition, the even parity and the odd parity bands (symmetric and antisymmetric combination of chiral basis) can have different $g$ factors. For a magnetic field along the $z$ direction, the generic Zeeman coupling has a form
\begin{equation}
H_Z=\Delta_{Z,+} \; \alpha_3 \gamma_5 +\Delta_{Z,-} \; \alpha_3 \gamma_5 \beta,
\end{equation} where $\Delta_{Z,\pm}$ respectively capture the sum and the difference between the Zeeman terms for the even and the odd parity bands. In the absence of the orbital coupling to the magnetic field, $\Delta_{Z,+}$ transforms the DSM into a WSM, whereas $\Delta_{Z,-}$ gives rise to a nodal ring. In the presence of the orbital coupling, the dispersion relations for $N=0$ levels are given by
\begin{eqnarray}
\epsilon_{N=0}(k_y,k_z,\alpha)&=& \mathrm{sgn}(\alpha)\bigg[t^2\sin^2 (k_zc)+\bigg \{ M(k_zc) \nonumber \\ &&+\frac{\hbar \omega_{c,\perp }}{2}+\Delta_{Z,-} \bigg \}^2\bigg]^{1/2}+\Delta_{Z,+} .
\end{eqnarray} The axial Zeeman coupling with $\Delta_{Z,+}\alpha_3 \gamma_5$, shifts the $N=0$ levels without opening a gap between them. In contrast, the antisymmetric part $\Delta_{Z,-}\alpha_3 \gamma_5 \beta$ tends to open a gap between the two $N=0$ levels. Therefore, within the $N=0$ subspace, the existence of a gap between the $\alpha=\pm 1$ levels is the most generic outcome. The Zeeman coupling also lifts the two fold spin degeneracy of the $N \neq 0$ Landau levels. However, the analytical expressions for the $N \neq 0$ Landau levels obtained from the solution of a quartic equation are tedious, and for this reason they are not reported here.

By restricting ourselves to the quantum limit, and away from the half-filling, where one of the $N=0$ levels (say $\alpha=+1$) is partially occupied, we can compute the LMC by following our general strategy. We use the following wavefunctions for the calculation of the matrix elements of impurity potential,
\begin{eqnarray}
\psi_{\alpha=+1}^T=\frac{1}{\sqrt{2}}\left[u(k_z), 0,0,v(k_z)\right] \varphi_0, \\
\psi_{\alpha=-1}^T=\frac{1}{\sqrt{2}}\left[-v(k_z), 0,0,u(k_z)\right] \varphi_0,
\end{eqnarray} where $\varphi_0$ is the wavefunction of the LLL for a three dimensional electron gas, and
\begin{eqnarray}
&&u^2(k_z)+v^2(k_z)=1, \\
&&u^2(k_z)-v^2(k_z)=\frac{t \sin k_z}{|\epsilon_{N=0}(k_y,k_z,\alpha)-\Delta_{Z,+}|}.
\end{eqnarray} These form factors modify the $|U_{eff}(2k_{F,0}(B))|^2$ obtained previously from Eq.~(\ref{ugs}) and Eq.~(\ref{ucs}) by a multiplicative factor
\begin{equation}
\frac{1}{4} [u^2(k_{F,0})-v^2(k_{F,0})]^2.
\end{equation} In addition we need to use $
k_{F,0}=2\pi^2 l^2_B \delta$, where $\delta$ is the carrier density, defined with respect to half-filling. The short range Gaussian impurities can again cause a $B$-linear positive LMC. In contrast, the ionic impurities cause a positive LMC varying as $\sim B^2$.

\subsection{Bi$_{1-x}$Sb$_x$}\label{bisb}

Recently, a negative LMR has been observed in Bi$_{1-x}$Sb$_{x}$.~\cite{kim1}  We believe this observed LMR is consistent with our predicted generic axial anomaly in the quantum limit. In particular,  the observed $B^2$ dependence of sigma(B) agrees with our predictions for the LMC in the quantum limit, when long-range ionic scattering is the dominant relaxation mechanism.  Given that the material is a semimetal which is closely related to a narrow gap semiconductor, it is expected that transport will be primarily determined by ionic impurity scattering. In addition our theory also provides a simple explanation for the upturn of the LMR beyond $B \sim 4T$ as arising from the residual short-range defect scattering in the system as shown in Fig.~\ref{LMR}.

Bi$_{0.96}$Sb$_{0.04}$ is a compensated semimetal, where the total number of carriers in the conduction band is equal to that in the valence band. There are three Dirac cones at the inequivalent L points. The Fermi energy intersects the conduction band around these Dirac points, producing three ellipsoidal electron pockets. These pockets are related by a rotation of $120^{\circ}$ about the trigonal axis. In addition, there is an ellipsoidal hole pocket at the T point, coming from the valence band, which can be modeled by a three dimensional parabolic electron gas. In the strong field limit, the positive LMR is mainly caused by the response of the hole pocket. The existing literature on magnetotransport and Nernst effect suggests that the quantum limit is achieved around $B \sim 3T$, when the B-field is directed along the trigonal axis~\cite{behnia2}. This happens due to the very low carrier density in the semimetallic Bi$_{1-x}$Sb$_{x}$. The density of holes is estimated to be $4 \times 10^{16} \; cm^{-3}$. An equal number of electrons are distributed among three electron pockets. A crude estimate using $l_B<n^{-1/3}$ leads to the threshold for the quantum limit $B_0 \sim 1T$, which compares very well with the experimentally determined $B_0$. Due to the presence of anisotropic mass and $g$-tensors, the quantum limit for some of the Fermi pockets can occur at a smaller $B_0$  when the magnetic field is applied along the bisectrix or the binary axes~\cite{behnia3}. These facts indeed corroborate our claim that the experiments are mostly being performed in the quantum limit. The upturn of the LMR beyond 4T is also in general agreement with our conclusions regarding the asymptotic behavior of the LMR, which is illustrated in Fig.~\ref{LMR}.

\subsection{ZrTe$_5$ and HfTe$_5$}\label{subsec:CME}In Ref.~\onlinecite{kharzeev}, a positive LMC varying as $B^2$ has been identified for ZrTe$_5$ up to $B=9 T$. For a very small field strength a positive LMR is observed, which has been ascribed to the weak antilocalization effects. In addition, the authors have performed ARPES measurements, which suggest the existence of linearly dispersing bands over a significant range of energy. At present there is some uncertainty whether the underlying Dirac fermion is truly massless or possesses a tiny mass. Again, given the small size of the Fermi pockets and the low carrier density, it is conceivable that the quantum limit is attained for a weak enough magnetic field. At least, the observed positive LMC varying as $B^2$ is consistent with our theory. Additional measurements are required to establish the threshold value of the field strength for the onset of the quantum limit. At least the previous magnetotransport measurements~\cite{levy1,levy2,izumi} suggest the existence of multiple Fermi pockets, and a carrier density for the hole pocket to $n_h \sim 6.5 \times 10^{23} m^{-3}$ for ZrTe$_5$, which will lead to the onset of quantum limit around a few Teslas. If the field is applied along the $b$ axis the threshold can be further decreased.

The authors of Ref.~\onlinecite{kharzeev} have assumed a constant relaxation rate for the axial charge, and invoked the existence of a chiral magnetic current for explaining the observed $B^2$ behavior of the LMC. The argument is as follows: (i) the steady state axial charge arising due to the ABJ anomaly, causes an axial chemical potential $\mu_5$, which is proportional to $\mathbf{E}\cdot \mathbf{B} \tau_b$, and (ii) there exists a chiral magnetic current
\begin{equation}
j_{CME}=\mu_5 B/(2\pi^2\hbar^2).
\end{equation}
This formula for the chiral magnetic current has been derived in the context of the unbounded linear dispersion of the Dirac fermions in Ref.~\onlinecite{fukushima}, and in this context it is a nondissipative, equilibrium current. For a bounded dispersion of solid state systems, there is no equilibrium chiral magnetic current~\cite{vazifeh,burkovsu}. This can be seen in the following way. A straight forward application of the formulas in Ref.~\onlinecite{fukushima} leads to
\begin{eqnarray}
&&j_{CME}=\sum_{N,\mathbf{k}}\langle \psi_N|\partial_{k_z} H|\psi_N \rangle \nonumber \\
&&=\frac{1}{2\pi l^2_B}\sum_n \int_{-\pi/c}^{\pi/c} \frac{dk_z}{2\pi} \frac{\partial \epsilon_N(k_z)}{\partial k_z} \theta(\mu-\epsilon_N(k_z)).\end{eqnarray} For an unbounded linear dispersion one sets $c \to 0$, and the LLL then represents two disconnected worlds of left and right handed fermions with different occupation numbers. The chiral magnetic current arises from these unbounded modes. For a bounded dispersion as we have shown for WSM and DSM, the left and the right handed worlds are connected through the bottom of the N=0 LLL (finite depth). Consequently, when we perform the integral over $k_z$ for a partially filled LL, we find the chiral magnetic current is proportional to $\epsilon_N(k_{F,N})-\epsilon_N(-k_{F,N})$, which always vanishes even in the presence of a $\mu_5 \neq 0$. As far as the completely filled LLs are concerned, their contribution vanishes due to the periodicity of integrand. From a calculation of diffusive magnetotransport of a WSM in a weak magnetic field, a positive LMC proportional to $B^2$ has been found in Ref.~\onlinecite{burkov1}, in agreement with the predictions of Ref.~\onlinecite{son} and Ref.~\onlinecite{kim1}. Since, the applied magnetic field strength up to 9 T is not a weak field regime, further experiments at higher field is needed to clarify various aspects.

Compared to ZrTe$_5$, HfTe$_5$ has a lower carrier density $n_h \sim 6 \times 10^{22} m^{-3}$~\cite{levy2,izumi}, which leads to a lower threshold for attaining quantum limit. Therefore, future measurements on HfTe$_5$ should also reveal the axial anomaly induced LMR in the quantum limit, which should display the generic behavior illustrated in Fig.~\ref{LMR}.

\section{Conclusions}\label{Conclusions}

In this manuscript we have shown that the existence of a finite LMC or LMR for any generic three dimensional metal is purely a quantum mechanical effect and is a direct consequence of the one dimensional axial anomaly of the dispersive LLs (see Eq.~(\ref{sigmab0}) and Eq.~(\ref{sigmabN})).
The occurrence of the axial anomaly is usually associated with the relativistic field theory in odd spatial dimensions, involving massless Dirac or Weyl fermions with unbounded linear dispersion. Therefore, one of our main findings that the axial anomaly is a generic feature of an arbitrary metal in parallel electric and magnetic fields may seem surprising and counterintuitive. Generally, the Adler Bell Jackiw equation is derived by evaluating a triangle diagram, by employing a Lorentz invariant regularization method. Such methods are well suited for the problems involving a fluctuating background gauge field, e.g., a non-abelian Yang Mills field, and the axial anomaly in that case is caused by the instanton density, which is proportional to $\mathrm{Tr}[\mathbf{E}\cdot \mathbf{B}]$. If we apply such methods of calculation for a system with a parabolic or a generic nonrelativistic dispersion, we will not find the axial anomaly.

However, the axial anomaly for the abelian background fields is somewhat subtle. For fluctuating electrodynamic fields, $\mathbf{E}$ and $\mathbf{B}$ are orthogonal and there is no axial anomaly. Only in the presence of externally imposed parallel electric and magnetic fields, the axial anomaly becomes finite. Since, a uniform background field strength is not a weak perturbation, many subtleties can often be missed out in a direct perturbative calculation performed in the plane wave basis. In the presence of a uniform external magnetic field, any three dimensional metal with conserved electric charge demonstrates Landau quantization of the cyclotron motion, and the dispersing Landau levels provide the appropriate description of the vacuum, and also serve as the proper basis for perturbative calculations. The Landau levels of a generic three dimensional system has one dimensional dispersion along $\mathbf{B}$ and an exact degeneracy factor $eB/h$. One dimensional conduction channels in the presence of an external electric field do possess charge pumping between the Fermi points (at the intersection of the one dimensional dispersion and the Fermi level), and this is described by Eq.~(\ref{ABJ1}). In a one dimensional system, this is related to the uniform acceleration of the center of mass in the presence of a uniform external electric field. This is the underlying reason for the emergence of the axial anomaly even for a three dimensional metal with parabolic dispersion. When $\mathbf{E}$ and $\mathbf{B}$ are parallel, the net charge pumping is obtained after multiplying both sides of Eq.~(\ref{ABJ1}).

Quite strikingly, we have shown that the appearance of a negative LMR is not in any way tied to the existence of an underlying Dirac or Weyl band structure. Rather it is intimately related to the type of scattering mechanism present in the system. We have calculated the generic axial anomaly induced LMC and LMR in the quantum limit in the presence of short-range neutral impurities and long-range ionic impurities. We have shown that ionic impurity scattering gives rise to a large positive LMC $\sigma \propto B^2$ in the quantum limit (see Eq.~\ref{ionicmagnetoconductivity}), while neutral short range impurities and point defects give rise to a non-monotonic dependence on the magnetic field (see Eq.~\ref{sigmab0short}), which is dependent on the underlying band structure. However, in the quantum limit the combined effects of neutral and ionic impurities initially lead to a negative LMR which ultimately becomes positive after passing through a minimum as demonstrated in Fig.~\ref{LMR} and Eq.~\ref{netLMR}.

We have also emphasized that due to the low density of carriers in gapless semimetals, the quantum limit can be reached for a moderate magnetic field.   As semiconductors generically possess low carrier densities, with large amounts of charged impurities (due to n and p type dopants), they are prime candidates to observe the axial anomaly induced negative LMR. We have shown qualitative agreement between our predicted LMR in the quantum limit (see Fig.~\ref{LMR} and Eq.~\ref{netLMR}) and the recent LMR measurements in TaAs, TaP, Na$_3$Bi, Bi$_{1-x}$Sb$_x$, ZrTe$_5$ and $\alpha$-(BEDT-TTF)$_2$I$_3$. The observed positive LMC in Bi$_{1-x}$Sb$_x$ and ZrTe$_5$ varying as $B^2$ up to a nonuniversal threshold is consistent with the LMC caused by long-range ionic impurity scattering. In particular, the quantum limit in Na$_3$Bi occurs around $B \sim 4T$, and the positive LMC varying as $B^2$ is observed up to $35 T$. Currently, this material provides the largest window for positive LMC in the quantum limit, and it lends strong support to our theoretical results regarding the importance of ionic scattering in gapless semiconductors. In addition, we predict that the observed upturn of the LMR is arising due to the short-range neutral impurity scattering. We have shown that the quantum limit in HfTe$_5$ occurs for a smaller magnetic field strength in comparison with ZrTe$_5$. Based on this we have predicted that the future LMR measurements on HfTe$_5$ will show pronounced evidence of generic axial anomaly in the quantum limit and the LMR will display the generic behavior consistent with Fig.~\ref{LMR} and Eq.~\ref{netLMR}).

We have shown that in layered materials with high carrier density, such that the underlying Fermi surface is a corrugated cylinder, one can obtain an effective situation like the quantum limit in a tilted magnetic field, when the tilt angle is in the vicinity of the Yamaji angle. Finally, we have shown qualitative agreement between our predictions and the recently observed negative LMR in PdCoO$_2$.

\appendix

\section{Boltzmann equation for magnetoconductivity}\label{boltzmannLMC}
Here we consider the Boltzmann equation approach for evaluating the transport lifetimes. We restrict ourselves to conventional scalar potential disorder, which do not mix the orbital or spin indices of the eigenstates.
The linearized Boltzmann equation for the transport lifetimes of different partially filled LLs is given by
\begin{eqnarray}
&&\sum_{N^\prime,k^\prime_y,k^\prime_z}W(N,k_y,k_z,\alpha;N^\prime,k^\prime_y,k^\prime_z,\alpha)\bigg[\tau_{tr,N,\alpha}(k_z)\nonumber \\&&- \frac{v_{N^\prime}(k^\prime_z)}{v_N(k_z)}\tau_{tr,N^\prime,\alpha}(k^\prime_z)\bigg]=1,\label{BMC1}
\end{eqnarray}where the impurity matrix element squared is given by
\begin{eqnarray}
&&W(N,k_y,k_z,\alpha;N^\prime,k^\prime_y,k^\prime_z,\alpha)=\frac{2\pi}{\hbar}n_i\int \frac{d^3q}{(2\pi)^3} |U(\mathbf{q})|^2 \nonumber \\ && \times |\langle N,k_y,k_z,\alpha|e^{i\mathbf{q}\cdot\mathbf{r}}|N^\prime,k^\prime_y,k^\prime_z,\alpha^\prime\rangle|^2 \delta(k_y-k^\prime_y+q_y)\nonumber \\
&& \times \delta(k_z-k^\prime_z+q_z)\delta(\epsilon_{N}(k_z)-\epsilon_{N^\prime}(k^\prime_z)).\label{BMC2}
\end{eqnarray} Here $|N,k_y,k_z,\alpha \rangle$ is the normalized wavefunction for the $N$ the LL and it can be a spinor. In addition, $\alpha$ represents the orbital or the spin degeneracy of the Landau levels.

For concreteness we will consider the LLs of the three dimensional electron gas. In this case,
\begin{eqnarray}
|\langle N,k_y,k_z,\alpha|e^{i\mathbf{q}\cdot\mathbf{r}}|N^\prime,k^\prime_y,k^\prime_z,\alpha^\prime\rangle|^2 = e^{-u}\; u^{|N^{'} - N|}\nonumber \\ \times |L^{|N^{'}-N|}_{\mathrm{min}(N,N^{'})}(u)|^2\frac{\Gamma \left(\mathrm{min}\{N,N^{'}\}+1\right)}{\Gamma \left(\mathrm{max}\{N,N^{'}\}+1\right)},\label{BMC3}
\end{eqnarray}where $u=q^2_\perp l^2_B/2$ and $L^{m}_{n}(x)$ is the associated Laguerre polynomial. The Eq.~(\ref{BMC1}) can rewritten in a more convenient form
\begin{eqnarray}
\frac{2\pi n_il^2_B}{\hbar}\sum_{N^\prime} \int \frac{dq_z}{2\pi} |U_{eff}(N,N^\prime ; q_z)|^2\bigg[\tau_{tr,N}(k_z)\nonumber \\ -\frac{v_{N^\prime}(k_z+q_z)}{v_N(k_z)}\tau_{tr,N^\prime}(k_z+q_z)\bigg]=1,
\end{eqnarray} where $U_{eff}(N,N^\prime ; q_z)$ are effective one dimensional scattering potentials obtained after integrating over $q_x$ and $q_y$. The expressions for these potentials are given by
\begin{eqnarray}
&&|U_{eff}(N,N^\prime ; q_z)|^2=\frac{\Gamma \left(\mathrm{min}\{N,N^{'}\}+1\right)}{\Gamma \left(\mathrm{max}\{N,N^{'}\}+1\right)}\int \frac{d^2 q_\perp}{(2\pi)^2l^2_B} \nonumber \\ && \times |U(q_\perp,q_z)|^2\exp\left(-\frac{q^2_\perp l^2_B}{2}\right)|L^{|N^{'}-N|}_{\mathrm{min}(N,N^{'})}\left(\frac{q^2_\perp l^2_B}{2}\right)|^2. \nonumber \\
\end{eqnarray} By setting $N=N^\prime$ we arrive at the general intra-LL scattering potential of Eq.~(\ref{UNqz}). The potential in Eq.~(\ref{U0qz}) is a special case, obtained by setting $N=N^\prime=0$ in Eq.~(\ref{UNqz}). For short range scattering due to neutral Gaussian impurities or zero range point impurities all the integrals can be performed analytically using
\begin{eqnarray}
&& \int^\infty_0 dx \; e^{-ax} \; x^\alpha \; |L^\alpha_\beta(x)|^2=\frac{\Gamma(\alpha+2\beta+1)}{\Gamma^2(\beta+1)}\frac{(a-1)^{2\beta}}{a^{\alpha+2\beta+1}} \nonumber \\
&&\times \; _2F_1\left(-\beta,-\beta;-\alpha-2\beta, \frac{a(a-2)}{(a-1)^2}\right).
\end{eqnarray} The integrals for Coulomb potential can also be performed analytically for small integers. Upon integrating over $q_z$ we obtain the following $M$ number of linear algebraic equations for the transport life times
\begin{widetext}
\begin{eqnarray}
\tau_{tr,N}(|k_{F,N}|)\bigg[\frac{U_{eff}(N,N;2k_{F,N})}{|k_{F,N}|}+\sum_{N^{'}\neq N}\frac{U_{eff}(N,N^{'};|k_{F,N}-k_{F,N^{'}}|)}{2|k_{F,N^{'}}|}+\sum_{N^{'}\neq N}\frac{U_{eff}(N,N^{'};|k_{F,N}+k_{F,N^{'}}|)}{2|k_{F,N^{'}}|}\bigg] \nonumber \\
+\sum_{N^{'} \neq N} \tau_{tr,N^{'}}(|k_{F,N^{'}}|) \bigg[\frac{U_{eff}(N,N^{'};|k_{F,N}-k_{F,N^{'}}|)}{2|k_{F,N}|}-\frac{U_{eff}(N,N^{'};|k_{F,N}+k_{F,N^{'}}|)}{2|k_{F,N}|}\bigg]=\frac{\hbar^3}{2mn_il^2_B},
\end{eqnarray}
\end{widetext}where $M$ is the number of partially occupied LLs. If we just consider the LLL, the transport lifetime becomes
\begin{eqnarray}
\tau_{tr,0}=\frac{\hbar^3 k_{F,0}}{2 m n_i l^2_B |U_{eff}(0,0,2k_{F,0})|^2} \nonumber \\
=\frac{\hbar^2 v_{F,0}}{2n_i l^2_B |U_{eff}(0,0,2k_{F,0})|^2}
\end{eqnarray} as has been discussed throughout the main text. The contributions of the short range scattering and the long range ionic scattering to the conductivity of a three dimensional electron gas in the absence of a magnetic field are respectively given by
\begin{widetext}
\begin{eqnarray}
&&\sigma^s(0)=\frac{ne^2}{m}\frac{2 \pi \hbar^3 a^3 k^2_F(0)}{ m n_s U_0^2} \left(\frac{\sqrt{\pi}}{4}\mathrm{Erf}(2 a k_F(0)) - a k_F(0)e^{-4 a^2 k^{2}_{F}(0)} \right)^{-1}, \\
&&\sigma^c(0)=\frac{ne^2}{m}\frac{\kappa^2 \hbar^3 k^3_F(0)}{2\pi m e^4 n_c}\left (\log \left[ 1+\frac{4k^2_F(0)}{q^2_{TF}(0)}\right]-\frac{4k^2_F(0)}{4k^2_F(0)+q^2_{TF}(0)}\right)^{-1},
\end{eqnarray}
\end{widetext}
which have been used in Fig.~\ref{fig:LMCelectrongas}. We have denoted the Fermi and the Thomas Fermi wavevectors at $B=0$ by $k_F(0)$ and $q_{TF}(0)$ respectively.

\section*{Acknowledgements} We acknowledge numerous valuable discussions with L. Balicas, N. Kikugawa, N. Hussey, V. Yakovenko, K. Behnia, B. Fauqu{\' e}, and L. Balents. This work is supported by JQI-NSF-PFC and LPS-CMTC.

\end{document}